\documentclass[aps,prd,nofootinbib,preprint,superscriptaddress]{revtex4}  
\usepackage{color}
\newcommand{\uu}{u_\uparrow}

\newcommand{\du}{d_\uparrow}

\newcommand{\tuu}{\tilde u_\uparrow}

\newcommand{\tdu}{\tilde d_\uparrow}

\begin{document}

\title{The permutation group $S_N$ and large $N_c$ excited baryons}

\author{Dan Pirjol} 
\affiliation{National Institute for Physics and Nuclear Engineering, 
Department of Particle Physics, 077125 Bucharest, Romania}

\author{Carlos Schat}
\affiliation{CONICET and Departamento de F\'{\i}sica, FCEyN, Universidad de Buenos Aires,
Ciudad Universitaria, Pab.1, (1428) Buenos Aires, Argentina} 
\affiliation{Departamento de F\'{\i}sica, Universidad de Murcia, E-30071 Murcia, Spain}

\date{\today}

\begin{abstract}
We study the excited baryon states for an arbitrary number of colors $N_c$ from
the perspective of the permutation group $S_N$ of $N$ objects. Classifying
the transformation properties of states and quark-quark interaction operators
under $S_N$ allows a general analysis of the spin-flavor structure of the 
mass operator of these states, in terms of a few unknown constants 
parameterizing the unknown spatial structure.
We explain how to perform the matching calculation of a general 
two-body quark-quark interaction onto the operators of the $1/N_c$ expansion. 
The inclusion of core and excited quark  operators is shown to be necessary. 
Considering the case of the negative parity $L=1$ states transforming in the $MS$ 
of $S_N$, we discuss the matching of the one-gluon and the Goldstone-boson exchange 
interactions.

\end{abstract}

\maketitle

\section{Introduction}

The nonrelativistic quark model (NRQM) \cite{De Rujula:1975ge} has been used with 
varying degree of success to describe the baryonic states, both for ground state 
and excited \cite{Isgur:1978xj} baryons.
Recently, the NRQM gained additional significance in the baryon sector, where it provides
an explicit realization of the contracted $SU(4)_c$ symmetry, which becomes manifest in the
large $N_c$ limit \cite{Dashen:1993jt}. 
This symmetry connects states with 
different spins and isospins, and provides an important organizing principle for the
construction of a systematic expansion in $1/N_c$ for physical quantities such as
masses and couplings. Many detailed applications to the physics of ground state
baryons, both heavy and light, have been presented.

The $1/N_c$ expansion has been applied also to orbitally excited baryons,
up to subleading order in $1/N_c$ 
\cite{Goity:1996hk,Pirjol:1997bp,Carlson:1998vx,PiSc,Schat:2001xr,Pirjol:2006ne,Matagne:2004pm}.
The structure of the operators contributing to the physical properties of
these states is more complicated. The construction of these operators makes use of the
decomposition of the spin-flavor states into ``core'' and ``excited'' quark subsystems.
The operators appearing in the $1/N_c$ expansion are the most general structures
which can be constructed from $SU(4)$ generators acting on the core and excited
quarks. 

The construction of the states and operators for excited states presented in 
Ref.~\cite{Carlson:1998vx} has been inspired by the quark model
picture of the excited states. To our knowledge a formal justification in terms 
of an underlying symmetry is still lacking. Such a symmetry argument would be
desirable for several reasons. First, it would help  in
order to establish the completeness of the set of operators contributing to any
given order in $1/N_c$. Second, such a symmetry argument could point the way for
the treatment of higher excitations with a more complicated spin-flavor structure.
Finally, in several recent papers \cite{Matagne:2006dj,Semay:2007cv} the validity of 
the usual approach based on the core+excited decomposition has been questioned, 
and a symmetry argument should settle the objections raised in these works.

In this paper we use the simple observation that the spin-flavor states of a 
system of $N$ quarks can be classified into irreps of $SU(4) \times S_N^{sf}$, with 
$S_N$ the permutation group of $N$ objects. Including also the orbital degrees of freedom, 
the complete permutation symmetry is $S_N \subset S_N^{orb} \times S_N^{sf}$,
the diagonal subgroup of the permutations acting on both orbital and spin-flavor
degrees of freedom. Of course, although $S_N$ is a good symmetry of the 
quark model Hamiltonian, $S_N^{sf}$ is not, and mixing between different
irreps can occur in general (configuration mixing).

The permutation symmetry $S_N^{sf}$ has nontrivial implications. 
It constrains the
form of the allowed operators which can contribute to any given matrix element.
The spin-flavor structure of the mass operator for these states contains only 
those spin-flavor operators ${\cal O}^{sf}$ which appear in
the decomposition of the interaction Hamiltonian ${\cal H} = 
\sum_{R} {\cal R}_R^{orb} {\cal O}_R^{sf} $ under  $S_N^{orb} \times S_N^{sf}$. 

The ground state baryons transform in the singlet irrep of $S_N^{sf}$, and thus
the only operators with nonzero matrix elements are also singlets of $S_N^{sf}$.
More interesting predictions are obtained for the orbitally excited baryons, which 
transform nontrivially under $S_N^{sf}$. We consider in some
detail the states transforming in the $[N-1,1]$ ($MS$) irrep of $S_N^{sf}$.

We perform the explicit decomposition of the most general two-body
quark-quark interaction into irreps of $S_N^{orb} \times S_N^{sf}$,
which gives a general expression for the mass operator of the excited states.
We present explicit results for the one-gluon exchange, and one Goldstone
boson exchange mediated quark-quark interaction. This generalizes
the $N_c=3$ results of Collins and Georgi \cite{Collins:1998ny} to 
arbitrary $N_c$ and also allows to extract the otherwise implicit $N_c$ 
dependence of the operators and their matrix elements. 

The analysis based on $S_N$ makes the connection to the $1/N_c$ expansion particularly
simple, allowing the matching of any microscopic quark-quark interaction onto the
operators of the $1/N_c$ expansion.
In addition to the class of operators transforming as $[N]$ and $[N-1,1]$ of 
$S_N^{sf}$, previously considered
in \cite{Collins:1998ny} for  $N_c=3$, we also discuss a new class of 
operators transforming in the $[N-2,2], [N-2,1,1]$ irreps, which appear only for 
$N_c > 3$. The matching calculation for two different models confirms 
the scalings and the completeness of the set of effective operators 
considered in \cite{Carlson:1998vx}. 
We find that the criticism of Matagne and Stancu \cite{Matagne:2006dj,Semay:2007cv}
of the usual $1/N_c$
expansion for excited baryons 
\cite{Goity:1996hk,Pirjol:1997bp,Carlson:1998vx,PiSc,Schat:2001xr,Pirjol:2006ne,Matagne:2004pm}
is unfounded, and the disagreement with existing work is due to ignoring 
operators transforming like non-singlets under $S_N^{sf}$.

As an application of our results, we point out that the
models for quark dynamics discussed here make definite predictions for the
phenomenology of the excited baryons, in the form of 
a hierarchy of the coefficients of operators in the $1/N_c$ expansion.
Comparing with the results of a fit we conclude that at leading order in $1/N_c$
the Goldstone boson exchange model is favored.

\section{The $S_N^{sf}$ permutation symmetry}
\label{Sec:Symm}

We describe in this section the $S_N$ transformation properties of the 
spin-flavor wave function of a system of $N=N_c$ quarks with spin and isospin.
For simplicity we consider only two light quark flavors $u, d$ and assume
isospin symmetry. 
In the nonrelativistic quark model, the spin-flavor wave functions appear
as building blocks for the total baryon wave function, along with the
orbital wave function. More importantly, they are relevant in the
context of the $1/N_c$ expansion, where the
quark representation provides a realization of the large $N_c$ spin-flavor 
symmetry \cite{Dashen:1993jt}.

The spin-flavor baryon states are constructed 
by taking products of one-body states with spin and isospin 
\begin{eqnarray}
| u_\uparrow\rangle\,, \quad
| u_\downarrow\rangle\,, \quad
| d_\uparrow\rangle\,, \quad
| d_\downarrow\rangle\,.
\end{eqnarray}
We keep track of the identity of the states of the $N_c$ quarks, 
such that in general, a state will not be left invariant under a permutation 
of the quarks. For any finite $N_c$, such a state will be completely specified
by giving its transformation under the symmetry group $SU(4) \times S_N$.

For any spin-flavor
state, the irreps of the spin-flavor $SU(4)$ and permutation group are related
such that they correspond to the same Young tableaux. 
For example, the
multiplicity of a state with given spin, isospin, and their projections
$|S,I;S_3, I_3 \rangle$ is given by the dimension of the $S_N$ irrep 
with the same Young diagram as the $SU(4)$ irrep.
For this reason it may
appear that the spin-flavor irrep is sufficient to describe the state. However,
it will be seen that the $S_N$ information can give constraints on the possible
form of the operators which do not follow from the spin-flavor structure alone.

We start by considering the ground state baryons. They transform in the 
completely symmetric irrep 
$S$ of $S_N^{sf}$. The only operators which 
have nonvanishing matrix elements between these states transform also 
in the $S$ irrep of the permutation group. 

Next consider states with nontrivial transformation under the permutation group.
As a first example of such states consider the nonstrange states in the 
$MS$ irrep of $S_N^{sf}$. An explicit analysis of the
spin-flavor structure in the
nonrelativistic quark model performed below in Sec.~\ref{Sec:NRQM} shows that
these states can be identified with the $L=1$ orbitally excited baryons.

Consider the mass operator of these states, and the constraints
imposed on it from the permutation group. It must transform like $R$ under the permutation
group, where $R$ is any irrep such that $MS \times R $ contains $MS$. The allowed
irreps must satisfy the following condition (see p.~257 of \cite{Hamm}): consider the 
Young diagrams 
of $R$ and of $MS$ (partition $[N-1,1]$). By removing one box from the $R$ diagram, 
and the addition of one box to the resulting diagram, the Young diagram of $MS$
must be obtained.
A simple examination of the respective Young diagrams shows that this condition is
satisfied by the following irreps of $S_N$ (see Appendix~A for notations)
\begin{eqnarray}\label{4irreps}
R = S , MS , E , A' \,.
\end{eqnarray}
The most general operator contributing to the mass operator of these states
will be thus a linear combination of operators transforming in each of the
irreps of $S_N^{sf}$ given in Eq.~(\ref{4irreps}).
The $S$ operators contain only building blocks
$S^i, T^a, G^{ia}$ for the entire hadron: total hadron spin, isospin and
the $G^{ia}$ operator. The $MS$ operators are constructed by combining operators
acting on the excited and core subsystems of the hadron consisting of 1 and
$N_c-1$ quarks respectively. Finally, the $E$ and $A'$ operators have a more
complicated structure, and have not been considered previously.

In the next section we will show by explicit calculation in the 
nonrelativistic quark model how each of these operators is generated from 
specific models of quark-quark interactions.
We consider in detail two such models, the one-gluon exchange 
and the Goldstone-boson exchange mediated interactions.

\section{$S_N$ and the nonrelativistic quark model}
\label{Sec:NRQM}

We consider in this section the spin-flavor structure of the orbitally excited
baryons transforming in the $MS$ irrep of $S_N^{sf}$ in the 
nonrelativistic quark model with $N_c$ quarks.
For $N_c=3$, these states correspond to the negative parity baryons with $L=1$ 
and transform in the $70^-$ of SU(6)  (or the $20^-$ of SU(4)).
This contains several spin-isospin multiplets of $SU(2) \times SU(3) \subset
SU(6)$. Denoting them as $^{2S+1}F$, these multiplets are $ {}^48 \oplus 
{}^28 \oplus {}^210 \oplus {}^21$. 

We would like to construct these states for arbitrary $N_c$ quarks, with 
special attention to their transformation properties under the permutation 
group. For generality we will keep $L$ arbitrary. Fermi 
statistics constrains the spin-flavor-orbital wave function to be completely
symmetric. Thus this wave function must transform under the
completely symmetric irrep of $S_N$, the permutation group acting on all
degrees of freedom of the $N$ quarks: orbital, spin and flavor. This group
is the diagonal subgroup of $S_N^{orb} \times S_N^{sf}$, the product of 
the permutation groups acting independently on the orbital and spin-flavor
degrees of freedom, respectively. Any symmetric state 
in orbital-spin-flavor space is decomposed into a sum of inner products of
irreps of $S_N^{orb} \times S_N^{sf}$ as
\begin{eqnarray}\label{Sdecomp}
S = \sum_{R= S, MS, \cdots} R^{orb} \otimes R^{sf} \,.
\end{eqnarray}
The sum is over states which transform in the same irrep $R$ under independent
permutations of the respective degrees of freedom.
We use here and in the following the notations of  Appendix~A for the irreps
of the permutation group.

The first term in the sum, with $R=S$, corresponds to states with orbital wave functions
symmetric under any exchange of two quarks. These states include the ground state
baryons, and some of their radial excitations. According to Eq.~(\ref{Sdecomp}), their spin-flavor wave 
function is also completely symmetric under permutations. For two light quark
flavors, this implies that they transform in the symmetric irrep of $SU(4)$.

\subsection{The $MS$ states and relation to CCGL}
\label{Sec:MSstates}

Consider next the second term in the sum Eq.~(\ref{Sdecomp}), corresponding to 
$R=MS$. They correspond
to orbitally excited baryons, for example the $L=1$ baryons in the $70^-$.  
A basis for the spin-flavor wave function can be constructed using the method of
the Young operators (see Appendix~A). This can be chosen as the set of $N-1$ 
wave functions, with $k=2, 3, \cdots, N$
\begin{eqnarray}\label{phidef}
|\phi_k \rangle &=& |q_k\rangle \otimes |i_c \rangle_{N-1} - 
|q_1\rangle \otimes |i_c\rangle_{N-1} \,,
\end{eqnarray}
where $|q_k\rangle$ denotes the spin-flavor state of the quark $k$, and 
$|i_c\rangle_{N-1}$ denotes the spin-flavor state of the subset of $N-1$ quarks (`core')
obtained by removing
quark $k$ from the $N$ quarks. The latter states are symmetric under any permutation
of the $N-1$ quarks. The states $\phi_i$ are not orthogonal, and have the scalar products 
\begin{eqnarray}\label{phinorm}
\langle \phi_i | \phi_j \rangle = S_{ij}\,, \qquad
S_{ij} = \left\{
\begin{array}{cc}
2 & , \quad i=j \\
1 & , \quad i\neq j \\
\end{array}
\right.
\end{eqnarray}

The basis states $\phi_k$ have the following transformations under the
action of the transpositions $P_{ij}$ (exchange of the quarks $i,j$)
\begin{eqnarray}\label{phitransp}
P_{1j} \phi_k &=& 
\left\{ 
\begin{array}{cc}
- \phi_k & , j=k \\
\phi_k - \phi_j & , j\neq k \\
\end{array}
\right. \\
P_{ij} \phi_k &=& \phi_k \qquad \mbox{ if } (i,j) \neq 1,k \\
P_{ik} \phi_k &=& P_{ki} \phi_k = \phi_i \quad  \mbox{ if } i\neq1 \,.
\label{phitransp2}
\end{eqnarray}
We will adopt these transformation relations as defining a basis for
the $MS$ irrep of $S_N$, both for states and operators.

A similar construction can be used also for the orbital wave functions with $MS$
permutation symmetry. They form a set of $N-1$ functions and will be 
denoted as $\chi_k^m(\vec r_1, \cdots , \vec r_N)$
with $k=2,3, \cdots, N$. The index $m=\pm 1, 0$ denotes 
the projection of the orbital angular momentum $\vec L$ along the $z$ axis.

For definiteness we adopt a Hartree representation for the orbital wave functions, in 
terms of one-body wave functions $\varphi_s(r)$ for the ground state orbitals, and
$\varphi_p^m(r)$ for the orbitally excited quark. 
The Young operator basis for the orbital wave functions is given by 
\begin{eqnarray}
\chi_k^m(\vec r_1, \cdots , \vec r_N) = 
\varphi_p^m(\vec r_k) \Pi_{i=1, i\neq k}^N \varphi_s(\vec r_i) -
\varphi_p^m(\vec r_1) \Pi_{i=2}^N \varphi_s(\vec r_i) \,.
\end{eqnarray}
These basis wave functions have the same transformation properties 
under transpositions
as the spin-flavor basis functions, Eqs.~(\ref{phitransp}-\ref{phitransp2}).
We emphasize that the main results of our paper do not depend on the use of the
Hartree representation, as explained in more detail below.

The complete spin-flavor-orbital wave function of a baryon $B$ with 
mixed-symmetric spin-flavor symmetry
is written as the $MS \times MS \to S$ inner product of the two basis wave 
functions, for the orbital and spin-flavor components, respectively.
Making explicit the spin-isospin degrees of freedom, the wave function is
given by
\begin{eqnarray}\label{MS}
|B;(LS)JJ_3,II_3\rangle = \sum_{m,S_3} \left(
\begin{array}{cc|c}
L & S & J \\
m & S_3 & J_3 \\
\end{array}
\right)
\sum_{k,l=2}^N \phi_k(SS_3,II_3) \chi_l^m M_{kl} \,.
\end{eqnarray}
For simplicity of notation, we will drop the explicit dependence of the
basis states on spin-isospin, and the SU(2) Clebsch-Gordan coefficient
coupling the orbital angular momentum with the quark spin $L+S=J$. 
They will be implicitly understood in all expressions written below. 
In Appendix~B we give an explicit example of this construction for a
baryon with $N_c=3$, and compare it with the baryon states constructed
in Ref.~\cite{Carlson:1998vx}.

The matrix of coefficients $M_{ij}$ are the Clebsch-Gordan
coefficients for the $MS \times MS \to S$ inner product of two irreps of $S_N$.
They can be determined by requiring that the state Eq.~(\ref{MS}) is left
invariant under the action of transpositions acting simultaneously on the
spin-flavor and orbital components. In the $MS$ basis defined by the
transformations Eqs.~(\ref{phitransp}) the matrix $M_{ij}$ is 
\begin{eqnarray}\label{Mmatrix}
\hat M = \left(
\begin{array}{ccccc}
1 & -\frac{1}{N-1} & -\frac{1}{N-1} & \cdots & -\frac{1}{N-1} \\
 -\frac{1}{N-1} & 1 &  -\frac{1}{N-1} & \cdots &  -\frac{1}{N-1} \\
 -\frac{1}{N-1} &  -\frac{1}{N-1} & 1 & \cdots &  -\frac{1}{N-1} \\
\cdots & \cdots & \cdots & \cdots & \cdots \\
 -\frac{1}{N-1} &  -\frac{1}{N-1} &  -\frac{1}{N-1} & \cdots & 1 \\
\end{array}
\right) \,.
\end{eqnarray}
Since 
any permutation can be represented as a product of transpositions, 
the state given in Eq.~(\ref{MS}) transforms indeed in the $S$ irrep of the 
overall $S_N$ group.

The form of the baryon state Eq.~(\ref{MS}) simplifies greatly with an
alternative choice for the basis of MS spin-flavor and orbital states.
This is the so-called Yamanouchi basis \cite{Hamm}, which is constructed such 
that i) the basis vectors are orthogonal and ii) each of them is invariant under 
permutations of mutually overlapping subsets of the $N$ quarks.
Considering the spin-flavor states, the Yamanouchi basis 
$\psi_k$, $k=2,\cdots , N$ is given by
\begin{eqnarray}
\psi_2 &=& \frac{1}{\sqrt{N(N-1)}}( \phi_2 + \phi_3 + \cdots + \phi_N ) \\
\psi_j &=& \frac{1}{\sqrt{(N-j+2)(N-j+1)}}
 ( -(N-j+1)\phi_{j-1} + \phi_j + \cdots + \phi_N ) \,,\quad j = 3, \cdots , N
\nonumber\,.
\end{eqnarray}
These states satisfy the normalization conditions $\langle \psi_i | \psi_j \rangle =
\delta_{ij}$. Also, using Eqs.~(\ref{phitransp}) one can see that 
$\psi_2,\psi_3, \psi_4 , \cdots , \psi_N$ are invariant
under permutations of the subsets $(2,3,\cdots ,N)\supset
(3, \cdots , N)\supset (4, \cdots , N) \supset
\cdots \supset (N-1,N)$ of the $N$ quarks, respectively.

Expressed in terms of the Yamanouchi  basis $\psi_k$, the matrix $M$,
giving the $MS\times MS \to S$ coupling, is proportional to the
unit matrix 
\begin{eqnarray}
M_{ij} = \frac{N}{N-1}\delta_{ij}\,,
\end{eqnarray}
such that the baryon state Eq.~(\ref{MS}) is given simply by (with $\xi_k$ the
Yamanouchi basis of orbital states)
\begin{eqnarray}\label{MSYam}
|B;(LS)JJ_3,II_3\rangle = \frac{N}{N-1}\sum_{m,S_3} \left(
\begin{array}{cc|c}
L & S & J \\
m & S_3 & J_3 \\
\end{array}
\right)
\sum_{k=2}^N \psi_k(SS_3,II_3) \xi_k^m  \,.
\end{eqnarray}

We pause at this point to compare the state Eq.~(\ref{MS}) with the $MS$ states 
constructed
in Ref.~\cite{Carlson:1998vx}, and commonly used in the literature in the 
context of the $1/N_c$ expansion. These states are constructed as tensor
products of an `excited' quark whose identity is fixed as quark $1$, with a 
symmetric `core' of $N-1$ quarks. We will refer to these states as CCGL states.
With this convention, the wave function used in Ref.~\cite{Carlson:1998vx}
has the form
\begin{eqnarray}
|CCGL \rangle = \Phi(SI) \,\, \varphi_p^m(\vec r_1) \Pi_{i=2}^N  
\varphi_s(\vec r_i) \,, \label{CCGLstate}
\end{eqnarray}
where $\Phi(SI)$ denotes the spin-flavor component, and the remainder is the 
orbital
wave function in Hartree form. By construction, the spin-flavor wave function
$\Phi(SI)$ transforms in the $MS$ irrep of $SU(4)$, and thus also like
$MS$ under $S_N$. It is symmetric under any exchange of the core quarks,
$P_{ij} \Phi(SI) = \Phi(SI)$ for $i,j\neq 1$. These two properties identify it uniquely 
in terms of the $MS$
basis functions defined in Eq.~(\ref{phidef}) as
$\Phi(SI) = \frac{1}{\sqrt{N(N-1)}}\sum_{k=2}^N \phi_k$. The normalization factor
is chosen such that the state is normalized as $\langle \Phi(SI)|\Phi(SI)\rangle 
= 1$.

Symmetrizing under the `excited' quark index $i=1,2, \cdots , N$, one finds
for the properly normalized symmetric state
\begin{eqnarray}
| CCGL \rangle \to \frac{1}{\sqrt{N}}\Sigma_{i=1}^N P_{1i} |CCGL\rangle = 
\frac{1}{N \sqrt{N-1}} \Sigma_{i=2}^N \Sigma_{j=2}^N (P_{1i} \phi_j) \chi_i  = 
- \frac{\sqrt{N-1}}{N} |B\rangle \,, \label{Bstate}
\end{eqnarray}
where the terms with different $i$ in the first sum are orthogonal states.
This gives the relation between the CCGL state and the 
symmetric state constructed above in Eq.~(\ref{MS}).

\subsection{$S_N\to S_N^{orb} \times S_N^{sf}$ decomposition of the interaction Hamiltonian}

In this Section we study the $S_N$ transformation properties of the
interaction Hamiltonian in the nonrelativistic quark model. For definiteness we 
adopt the one-gluon exchange potential,
which follows from a perturbative expansion in $\alpha_s(m_Q)$ in the heavy quark limit
$m_Q \gg \Lambda_{QCD}$. We formulate our discussion in sufficiently general terms 
to allow the treatment of any other Hamiltonian containing only two-body 
interactions.

We consider a Hamiltonian containing a spin-flavor symmetric term $H_0$ 
(the confining potential and kinetic terms), plus spin-isospin dependent 
two-body interaction terms $V_{ij}$ 
\begin{eqnarray}\label{Hamilton}
H = H_0 + g_s^2 \sum_{i<j} \frac{\lambda_i^a}{2} \frac{\lambda_j^a}{2} V_{ij} \to
H_0 - g_s^2\frac{N_c+1}{2N_c}\sum_{i<j} V_{ij} \,,
\end{eqnarray}
where $\lambda^a$ are the generators of SU(3) color in the fundamental 
representation, and $g_s$ is the strong gluon coupling to the quarks, scaling 
like $g_s\sim O(N_c^{-1/2})$.
The second equality holds on color-singlet hadronic states, on which
the color interaction evaluates to the color factor 
$(N_c+1)/(2N_c)$.
We will restrict ourselves only to color neutral hadronic states in this paper.

In the nonrelativistic limit, the two-body interaction $V_{ij}$ contains three 
terms: the spin-spin interaction $(V_{ss})$, the quadrupole interaction 
$V_q$ and the spin-orbit terms $V_{so}$. 
We write these interaction terms in a slightly more general form as 
\cite{Collins:1998ny}, where $f_{0,1,2}(r_{ij})$ are unspecified functions
of the interquark distances
\begin{eqnarray}
V_{ss} &=& \sum_{i < j=1}^N f_0(r_{ij}) \vec s_i \cdot \vec s_j  \,, \\
V_{q} &=& \sum_{i < j=1}^N f_2(r_{ij}) \Big[3(\hat r_{ij} \cdot \vec s_i)(\hat r_{ij} \cdot \vec s_j)
 -  (\vec s_i \cdot \vec s_j) \Big]  \,, \\
V_{so} &=& \sum_{i < j=1}^N f_1(r_{ij}) \Big[ 
(\vec r_{ij} \times \vec p_i ) \cdot \vec s_i - (\vec r_{ij} \times \vec p_j ) \cdot \vec s_j \\
& & \hspace{1.5cm}+ 2 (\vec r_{ij} \times \vec p_i ) \cdot \vec s_j - 2 
(\vec r_{ij} \times \vec p_j ) \cdot \vec s_i \Big]  \,. \nonumber
\end{eqnarray}
 
We would like to decompose the Hamiltonian $H$ into a sum of terms 
transforming according to irreps of the permutation 
group acting on the spin-flavor degrees of freedom $S_N^{sf}$. 
The operators in Eq.~(\ref{Hamilton}) are two-body interactions, of the
generic form
\begin{eqnarray}
V = \sum_{1\leq i<j\leq N} {\cal R}_{ij} {\cal O}_{ij} \,,
\end{eqnarray}
where ${\cal R}_{ij}$ acts only on the orbital coordinates of the quarks 
$i,j$, and
${\cal O}_{ij}$ acts only on their spin-flavor degrees of freedom. For 
example, the
spin-spin interaction $V_{ss}$ has ${\cal R}_{ij} = f_0(r_{ij})$ and 
${\cal O}_{ij} = \vec s_i \cdot \vec s_j$.
Of course, $V$ must be symmetric under
any permutation of the $N$ quarks, but the transformation of
the spin-flavor and orbital factors separately can be more complicated.
We distinguish two possibilities for the transformation of these 
operators under a transposition of the 
quarks $i,j$, corresponding to the two irreps of $S_2$:

i) symmetric two-body operators:
$P_{ij} {\cal R}^s_{ij} P^{-1}_{ij} = {\cal R}^s_{ij}$ and
$P_{ij} {\cal O}^s_{ij} P^{-1}_{ij} = {\cal O}^s_{ij}$.

ii) antisymmetric two-body operators: 
$P_{ij} {\cal R}^a_{ij}P^{-1}_{ij} = -{\cal R}^a_{ij}$ and
$P_{ij} {\cal O}^a_{ij}P^{-1}_{ij} = -{\cal O}^a_{ij}$.

The spin-spin and quadrupole 
interactions $V_{ss}, V_q$ are composed of symmetric two-body
operators, while the spin-orbit interaction $V_{so}$ contains both symmetric and 
antisymmetric components.

In general, the $k-$body operators can be classified into irreps of the
permutation group of $k$ objects $S_k$. For example, there are three classes
of 3-body operators, corresponding to the $S,MS,A$ irreps of $S_3$.

We start by considering the symmetric two-body operators.
The set of all spin-flavor operators ${\cal O}^s_{ij}$ (and analogous for 
the orbital operators ${\cal R}^s_{ij}$) 
with $1 \leq i < j \leq N$ 
form a $\frac12 N(N-1)$ dimensional
reducible representation of the $S_N$ group, which contains the following irreps
\begin{eqnarray}
\{ {\cal O}^s_{ij} \} = S \oplus MS \oplus E \,.
\end{eqnarray}  
We will use as a basis for the operators on the right-hand side the Young 
operator basis supplemented by the phase convention Eq.~(\ref{phitransp})
as explained in Appendix~A. The projection of the operators onto irreps of
$S_N$ are as follows.

\underline{The $S$ projection.}
\begin{eqnarray}
{\cal O}^S = \sum_{i<j} {\cal O}^s_{ij}
\end{eqnarray}

\underline{The $MS$ projection.} 
\begin{eqnarray}
{\cal O}^{MS}_k = \sum_{2\leq j \neq k\leq N} ( {\cal O}^s_{1j} - {\cal O}^s_{kj} )\,,
\qquad k = 2, 3, \cdots , N
\end{eqnarray}

\underline{The $E$ projection.}
These operators can be labeled by the three integers $i,j,k$ appearing in the
standard Young tableau $[1i\cdots ][jk]$ corresponding to the respective
operator, and are given by 
\begin{eqnarray}
{\cal O}^{E}_{ijk} = {\cal O}^s_{1i} - {\cal O}^s_{1k} - {\cal O}^s_{ij} +
{\cal O}^s_{jk}\,,
\qquad k > i,j= 2, \cdots , N-1 \,.
\end{eqnarray}
In Appendix~A we give as an illustration the complete basis of the $S,MS,E$
operators for $N_c=5$.

The interaction $V^{symm}$ constructed with symmetric two-body operators 
is symmetric under $S_N$, and its 
decomposition under $S_N \subset S_N^{orb} \times S_N^{sf}$ has the form
\begin{eqnarray}\label{Oeven}
V^{symm} = \frac{2}{N(N-1)} {\cal R}^S {\cal O}^S + 
\frac{N-1}{N(N-2)} \sum_{j,k=2}^N  {\cal R}^{MS}_j {\cal O}^{MS}_k M_{jk} +  
c_{E}^{symm} \sum_{j,k=1}^{\frac12 N(N-3)} {\cal R}^E_j {\cal O}^E_k N_{jk}  \,.
\nonumber \\
\end{eqnarray}
The matrices $M_{jk}$ and $N_{jk}$ are Clebsch-Gordan coefficients for the
$MS \times MS \to S$ and $E \times E \to S$ reductions for the irreps of the $S_N$
group. The matrix $M_{jk}$ is given in Eq.~(\ref{Mmatrix}) in explicit form.
We do not have explicit results for $c_E$ and the matrix $N_{jk}$ for arbitrary
$N$, although for any given $N$ they can be found as explained in Appendix~A.

We consider next the case of the antisymmetric two-body operators ${\cal O}^a_{ij}$. They
form a $\frac12 N(N-1)$ dimensional reducible representation of $S_N$, which is 
decomposed into irreps as
\begin{eqnarray}
\{ {\cal O}^a_{ij} \} = MS \oplus A' \,.
\end{eqnarray}  
This can be verified by noting that the sum of the dimensions of the irreps
on the right-hand side is indeed $\frac12 N(N-1)$. We list in Appendix~A the explicit
basis of operators for the $A'$ irrep with $N=5$.
The projection of these operators onto irreps of $S_N$ gives the following 
basis operators.

\underline{The $MS$ projection. }
\begin{eqnarray}
{\cal O}^{MS}_k = \sum_{j=1}^N ( {\cal O}^a_{1j} - {\cal O}^a_{kj} )\,,
\qquad k = 2, 3, \cdots , N
\end{eqnarray}

\underline{The $A'$ projection.}

These operators can be labelled by the two integers $j,k$
appearing in the first column 
of the standard Young tableaux $[1...][j][k]$ corresponding to the given operator. 
They can be chosen as
\begin{eqnarray}\label{OA}
{\cal O}^{A'}_{jk} = 2 {\cal O}^a_{1j} + 2 {\cal O}^a_{jk} - 2 {\cal O}^a_{1k}\,,
\qquad 2 \leq j < k \leq N
\end{eqnarray}

The decomposition of the 
interaction $V^{anti} = \sum_{i<j} {\cal R}^a_{ij} {\cal O}^a_{ij} $ consisting of
products of antisymmetric two-body interactions, into operators transforming
as irreps of $S_N^{orb} \times S_N^{sf}$ has a similar form to 
Eq.~(\ref{Oeven})
\begin{eqnarray}\label{Oodd}
V^{anti} = 
\frac{N-1}{N^2} \sum_{j,k=2}^N  {\cal R}^{MS}_j {\cal O}^{MS}_k M_{jk} +  
c^{anti}_{E} \sum_{j,k=1}^{\frac12 (N-1)(N-2)} 
{\cal R}^{A'}_j {\cal O}^{A'}_k Q_{jk}  \,,
\end{eqnarray}
where $M_{jk}$ is given in Eq.~(\ref{Mmatrix}), and $Q_{jk}$ is the Clebsch-Gordan
coefficient for the reduction $A' \times A' \to S$.

\subsection{Mass operator - One-gluon exchange interaction}

The discussion of the interaction Hamiltonian in the previous section was completely
general, and did not assume anything about the hadronic states. In this section 
we consider its matrix elements on the $|B\rangle$ states constructed
in Sec.~\ref{Sec:MSstates}, with special regard to their spin-flavor structure. 
The permutation symmetry reduces the matrix elements of the orbital operators
${\cal R}_{ij}$ to a small number of unknown reduced
matrix elements, which can be expressed as overlap integrals. Counting the
number of the contributing irreps gives that there are 3 undetermined reduced matrix
elements for each symmetric two-body operator, and 2 reduced matrix elements 
for each antisymmetric two-body operator.

We compute the matrix elements of the 2-body interaction Hamiltonian
on the $|B\rangle $ states constructed in Eq.~(\ref{MS}). 
The matrix elements of the spin-flavor operators on the basis functions $\phi_i$
are given by 
\begin{eqnarray}\label{WEOS}
\langle \phi_i | {\cal O}^S |\phi_j \rangle &=& 
\langle {\cal O}^S \rangle (1 + \delta_{ij}) \,, \\
\label{WEOMS}
\langle \phi_i | {\cal O}^{MS}_k |\phi_j \rangle &=& 
\langle {\cal O}^{MS} \rangle (1 - \delta_{ik}\delta_{ij}) \,,\\
\label{WEOE}
\langle \phi_i | {\cal O}^E_{klm} |\phi_j \rangle &=& 
\langle {\cal O}^E \rangle
\frac12 [(-\delta_{ik}+\delta_{im})(1+\delta_{jl}) + 
(-\delta_{jk}+\delta_{jm})(1+\delta_{il})] \,,
\end{eqnarray}
where 
$\langle {\cal O}^S\rangle, \langle {\cal O}^{MS}\rangle, \langle {\cal O}^{E}\rangle$ 
are reduced matrix elements.
The proportionality of the matrix elements to just one reduced
matrix element follows from the Wigner-Eckart theorem for the
$S_N$ group. The form of the Clebsch-Gordan coefficients is specific
to the $MS$ basis used in this paper, and can be derived by repeated 
application of Eqs.~(\ref{phitransp}) to the states and operators.

For the orbital operators ${\cal R}$, an additional complexity is 
introduced by the presence of the magnetic quantum numbers of the 
initial and final state orbital basis functions $\chi_p^m$. The 
dependence on $m,m'$ is given by the Lorentz structure of the orbital
operator. The simplest case corresponds to a Lorentz scalar,
for which the matrix elements are given by
\begin{eqnarray}
\langle \chi_i^{m'} | {\cal R}^S |\chi_j^m \rangle &=& 
\langle {\cal R}^S \rangle (1+\delta_{ij})  \; \delta_{mm'}  \,, \\
\label{RMS}
\langle \chi_i^{m'} | {\cal R}^{MS}_k |\chi_j^m \rangle &=& 
\langle {\cal R}^{MS} \rangle  (1 - \delta_{ik}\delta_{ij}) \;  \delta_{mm'} \,,\\
\langle \chi_i^{m'} | {\cal R}^E_{kln} |\chi_j^m \rangle &=& 
\langle {\cal R}^E \rangle
\frac12 [(-\delta_{ik}+\delta_{in})(1+\delta_{jl}) + 
(-\delta_{jk}+\delta_{jn})(1+\delta_{il})] \; \delta_{mm'}\,,
\end{eqnarray}

Inserting these expressions into Eq.~(\ref{Oeven}) and combining all factors 
we find the following result for the matrix element of a symmetric two-body 
operator $V^{symm}$, 
expressed as a sum over irreps of $S_N^{\rm orb}\times
S_N^{\rm sp-fl}$
\begin{eqnarray}\label{symm2body}
\langle V^{symm}  \rangle \equiv &&
\frac{\langle B | V^{symm} | B \rangle}{\langle B| B \rangle}\\
= && \frac{2}{N(N-1)}
\langle {\cal R}^S \rangle \langle {\cal O}^S \rangle
+ \frac{1}{N}
\langle {\cal R}^{MS} \rangle \langle {\cal O}^{MS} \rangle 
+ \frac{N(N-3)}{4(N-1)} \langle {\cal R}^{E} \rangle 
\langle {\cal O}^{E}\rangle \,.\nonumber
\end{eqnarray}
The contribution of the $E$ operators in the last term is obtained using an
alternative method, sketched in Appendix C, which does not require the knowledge 
of the Clebsch-Gordan coefficients $N_{jk}$.

The reduced matrix elements depend on the precise form of the
interaction. We consider for definiteness the spin-spin interaction $V_{ss}$
in some detail. For this case the reduced matrix elements of the symmetric operators
are given by
\begin{eqnarray}
\langle {\cal R}^S \rangle &=& 
\frac12 (N-1)(N-2) {\cal I}_{s} + (N-1) {\cal I}_{\rm dir} - {\cal I}_{\rm exc} \,,\\
\langle {\cal O}^S \rangle &=& \langle \Phi(SI) |\frac12 \vec S\,^2 -
\frac38 N |\Phi(SI)\rangle  \,.
\end{eqnarray}
For the ease of comparison with the literature on the $1/N_c$ expansion for 
excited baryons, we expressed the reduced matrix element of the spin-flavor
operator as a matrix element on  the state $|\Phi(SI)\rangle$ where the
excited quark is quark no. 1. This state is given by 
$\Phi(SI) = \frac{1}{\sqrt{N(N-1)}}\sum_{k=2}^N \phi_k$.
The reduced matrix elements of the
orbital operator are expressed in terms of the 3 overlap integrals over the
one-body wave functions
\begin{eqnarray}\label{overlap}
{\cal I}_s &=& \int d\vec r_1 d\vec r_2 f_0(r_{12}) |\varphi_s(\vec r_1)|^2 
|\varphi_s(\vec r_2)|^2 \,, \\
{\cal I}_{\rm dir} &=& \int d\vec r_1 d\vec r_2 f_0(r_{12}) 
|\varphi_s(\vec r_1)|^2 |\varphi_p^m(\vec r_2)|^2 
\nonumber  \,, \\
{\cal I}_{\rm exc} &=& \int d\vec r_1 d\vec r_2 f_0(r_{12}) 
\varphi_s^*(\vec r_1) \varphi_p^{m*}(\vec r_2) 
 \varphi_p^{m}(\vec r_1) \varphi_s (\vec r_2) \,. \nonumber
\end{eqnarray}

The spin-flavor operator transforming in the $MS$ irrep is ${\cal O}_k^{MS} =
(\vec s_1 - \vec s_k) \cdot \vec S$, and the corresponding orbital operator
is ${\cal R}_k^{MS} = \sum_{j=2, j\neq k}^N [f_0(r_{1j}) - f_0(r_{kj})]$.
Their reduced matrix elements are
\begin{eqnarray}
\langle {\cal R}^{MS} \rangle &=& (N-2) 
({\cal I}_{\rm dir} - {\cal I}_{s}) - 2 {\cal I}_{\rm exc} \,, \\
\langle {\cal O}^{MS} \rangle &=& \frac{1}{N-2}
\langle \Phi(SI) |- \vec S\,^2
+ N \vec s_1 \cdot \vec S_c + \frac34 N |\Phi(SI)\rangle  \,,
\label{OMS}
\end{eqnarray}
where the overlap integrals are the same as defined in Eqs.~(\ref{overlap}). We
denoted $\vec S_c$ the `core' spin, defined as $\vec S_c = \vec S - \vec s_1$. 
The reduced matrix element of the orbital operator $\langle {\cal R}^{MS} \rangle$
is computed by taking representative values of $i,j,k$ in Eq.~(\ref{RMS}) and
evaluating the integrals. The derivation of the spin-flavor reduced matrix
element $\langle {\cal O}^{MS} \rangle$ is given in Appendix~D.
We postpone the discussion of the $E$ operators for a later section, 
as they require a separate treatment.

The quadrupole interaction is treated in a similar way. This is written
as
\begin{eqnarray}
V_{q} = \frac12 \sum_{i<j} f_2(r_{ij}) (3 \hat r_{ij}^a \hat r_{ij}^b  -
\delta^{ab}) (s_i^a s_j^b + s_i^b s_j^a) \equiv 
\sum_{i<j} {\cal Q}^{ab}_{ij} {\cal O}^{ab}_{ij} \,,
\end{eqnarray}
where ${\cal Q}_{ij}^{ab}$ acts only on the orbital degrees of freedom, 
${\cal O}_{ij}^{ab}=s_i^a s_j^b + s_i^b s_j^a$ on spin-flavor, and 
$a,b$ are spatial indices.
As mentioned, this is a symmetric two-body operator. The projections onto $S$ and 
$MS$ of each of the factors are obtained as explained above. The symmetric 
projections are
\begin{eqnarray}
{\cal Q}_{S}^{ab} &=& \sum_{i<j}  {\cal Q}^{ab}_{ij} = \frac12 
\sum_{i<j} f_2(r_{ij}) (3 \hat r_{ij}^a \hat r_{ij}^b  -
\delta^{ab})  \,, \\
{\cal O}^{ab}_S &=& \sum_{i<j}  {\cal O}^{ab}_{ij} = \frac12 
\Big( \{S^a\,,  S^b\} - \frac12 N \delta^{ab} \Big) \,,
\end{eqnarray}
where the braces denote symmetrization with respect to the spatial indices 
$a,b$, $\{ X^a, Y^b\} = X^a Y^b + X^b Y^a$. 
Their matrix elements between the $MS$ basis for orbital and spin-flavor 
states are given by the Wigner-Eckart theorem for the permutation group
\begin{eqnarray}
\langle \chi_i^{m'} |{\cal Q}_S^{ab} |\chi_j^m \rangle &=& S_{ij} 
\Big( \frac12 \{ L^a\,, L^b\} - \frac13 L(L+1)\delta^{ab}\Big)_{m' m} 
\langle {\cal Q}_S\rangle \,, \\
\langle \phi_i |{\cal O}_S^{ab} |\phi_j \rangle &=& S_{ij}  
\langle{\cal O}_S^{ab}\rangle \,.
\end{eqnarray}
The dependence on the magnetic quantum numbers $m,m'$ in the orbital 
matrix element 
is given by the most general symmetric and traceless tensor which can 
be formed from
the angular momentum. We will denote it as 
$L_2^{ab}= \frac12 \{L^a ,  L^b\} - \frac13 L(L+1)\delta^{ab}$.

The reduced matrix elements are given by
\begin{eqnarray}
& &\langle {\cal Q}_S\rangle = (N-1) {\cal K}_{\rm dir} - {\cal K}_{\rm exc} \,, \\
& &\langle {\cal O}_S^{ab}\rangle = \frac12\langle \Phi(SI) |
\{S^a\,, S^b\} - \frac12 N \delta^{ab} |\Phi(SI)\rangle \,.
\end{eqnarray}
The convolution integrals ${\cal K}_{\rm dir, exc}$ appearing in the reduced
matrix element of the orbital operator ${\cal Q}_S$ are similar to those
introduced for the spin-spin interaction. 
\begin{eqnarray}
& & {\cal K}_{\rm dir} \Big(L_2^{ab}\Big)_{m' m} \equiv \frac12 
\int d\vec r_1 d \vec r_2 f_2(r_{12}) (3 \hat r_{12}^a \hat r_{12}^b - \delta^{ab} )
|\varphi_s(\vec r_1)|^2 \varphi_p^{m' *}(\vec r_2) \varphi_p^{m} (\vec r_2)  \,, \\
& & {\cal K}_{\rm exc} \Big(L_2^{ab}\Big)_{m' m} \equiv \frac12
\int d\vec r_1 d\vec r_2 f_2(r_{12}) (3 \hat r_{12}^a \hat r_{12}^b - \delta^{ab} )
\varphi_s^*(\vec r_1) \varphi_p^{m' *}(\vec r_2 ) \varphi_p^{m} (\vec r_1) 
\varphi_s(\vec r_2)  \,.
\end{eqnarray}

The $MS$ projections of the operators are
\begin{eqnarray}
[ {\cal Q}^{ab} ]^{MS}_k &=& \frac12
\sum_{j=2\neq k}^N f_2(r_{1j}) (3 \hat r_{1j}^a \hat r_{1j}^b - \delta^{ab}) -
\frac12 
\sum_{j=2\neq k}^N f_2(r_{kj}) (3 \hat r_{kj}^a \hat r_{kj}^b - \delta^{ab}) \,, \\ 
 {[} {\cal O}^{ab} ]^{MS}_k &=& \{ s_1^a - s_k^a, S^b\}  \,.
\end{eqnarray}
Their reduced matrix elements are given by
\begin{eqnarray}
\langle {\cal Q}_{MS} \rangle &=& 
(N-2) {\cal K}_{\rm dir} - 2 {\cal K}_{\rm exc}  \,, \\
\langle {\cal O}_{MS}^{ab} \rangle &=& 
\frac{1}{N-2} \langle \Phi(SI) | N \{ S_c^a, s_1^b \} - \{S^a\,, S^b\}
+ \frac12 N \delta^{ab} |\Phi(SI) \rangle \,.
\end{eqnarray}

Finally we discuss the computation of the matrix element of the 
spin-orbit interaction $V_{so}$, which
is more involved. As mentioned, this operator can be written as a sum of 
symmetric and antisymmetric two-body operators
\begin{eqnarray}
V_{so} = V_{so}^{symm} + V_{so}^{anti} =
\frac32 \sum_{i < j} \vec {\cal L}_{ij}^s \cdot (\vec s_i + \vec s_j) -
\frac12 \sum_{i < j} \vec {\cal L}_{ij}^a \cdot (\vec s_i - \vec s_j) \,,
\end{eqnarray}
where we defined the symmetric and antisymmetric orbital operators
\begin{eqnarray}
\vec {\cal L}_{ij}^s = f_1(r_{ij}) \vec r_{ij} \times (\vec p_i - \vec p_j)\,,\qquad
\vec {\cal L}_{ij}^a = f_1(r_{ij})\vec r_{ij} \times (\vec p_i + \vec p_j)\,.
\end{eqnarray}
They are decomposed as $S\oplus MS\oplus E$ and $MS\oplus A'$ respectively, and thus their 
matrix elements taken between $MS$ states of $S_N$ acting on spin-flavor
depend on 5 reduced matrix elements. At $N_c=3$ the $E$ operators do not exist,
and thus this number reduces to 4, which agrees with the 
counting of Ref.~\cite{Collins:1998ny}. Table \ref{table_redme} summarizes the 
counting of the independent reduced matrix elements arising from each of the
terms in the quark interaction Hamiltonian.

Consider first the symmetric piece of the spin-orbit interaction 
$V_{so}^{symm}$. The projections
of the orbital and spin-flavor factors onto $S$ and $MS$ irreps of the $S_N$ 
group are: the symmetric $S$ components
\begin{eqnarray}
& &[\vec {\cal L}^s]_S = \sum_{i < j}  \vec {\cal L}^s_{ij} \,, \\
& &[\vec s_i + \vec s_j]_S = (N-1) \vec S \nonumber
\end{eqnarray}
and the $MS$ components
\begin{eqnarray}
& &[\vec {\cal L}^s]_k^{MS} = \sum_{j=2\neq k}^N 
 \vec {\cal L}^s_{1j} - \sum_{j=2\neq k}^N 
 \vec {\cal L}^s_{kj} \,, \\
& &[\vec s_i + \vec s_j]^{MS}_k = (N-2) (\vec s_1 - \vec s_k) \,. \nonumber
\end{eqnarray}

The matrix element of the $V_{so}^{symm}$ operator
is given by the same relation as Eq.~(\ref{symm2body}) except that
the reduced matrix elements are vectors. Taking into account that the
only available vector is the angular momentum $\vec L$, the most 
general form for the matrix elements of the orbital operators
$\vec {\cal L}_{ij}^s$ are given as
\begin{eqnarray}
\langle \chi_i^{m'} | [\vec {\cal L}^s]_S | \chi_j^m \rangle &=& S_{ij} 
(\vec L)_{m'm} \langle {\cal L}_S^s \rangle  \,, \\
\langle \chi_i^{m'} | [\vec {\cal L}^s]_k^{MS} | \chi_j^m \rangle &=&
(1 - \delta_{ik} \delta_{jk}) 
(\vec L)_{m'm} \langle {\cal L}_{MS}^s \rangle  \,.
\end{eqnarray}

Collecting together all factors, the matrix element of the $V_{so}^{symm}$
interaction is given by (we neglect here the contribution of the $E \otimes E$ 
operators to the matrix element, which does not contribute for $N_c=3$)
\begin{eqnarray}\label{Vsosymm}
\frac{\langle B | V_{so}^{symm} | B \rangle}{\langle B | B \rangle} = 
\frac{2}{N}\vec L \cdot \vec S 
\langle {\cal L}_S^s \rangle + \frac{1}{N}
\Big( N \vec s_1 \cdot \vec L - \vec S \cdot \vec L\Big)
\langle {\cal L}_{MS}^s \rangle \,.
\end{eqnarray}
As usual, the expression on the rhs is understood as an operator acting on the CCGL 
type spin-flavor states $\Phi(SI)$, for which the excited quark is quark no.1.

The scalar coefficients appearing as reduced matrix elements of the $S$ and $MS$
orbital operators are  given by overlap integrals over the two-body interaction
\begin{eqnarray}
\langle {\cal L}_S^s \rangle &=& (N-1) {\cal J}_{\rm dir}^s - {\cal J}_{\rm exc}^s \,, \\
\langle {\cal L}_{MS}^s \rangle &=&
(N-2) {\cal J}_{\rm dir}^s - 2 {\cal J}_{\rm exc}^s  \,, \nonumber
\end{eqnarray}
where the coefficients ${\cal J}^s_{\rm dir, exc}$ are defined by the 
integrals 
\begin{eqnarray}
(\vec L)_{m'm} {\cal J}_{\rm dir}^s &=&
\int d\vec r_1 d\vec r_2 
f_1(r_{12}) \vec r_{12} \times (\vec p_1 - \vec p_2) 
|\varphi_s(\vec r_1)|^2 \varphi_p^{m' *}(\vec r_2) \varphi_p^{m} (\vec r_2) \,,  \\
(\vec L)_{m'm} {\cal J}_{\rm exc}^s & = &
\int d\vec r_1 d\vec r_2 
f_1(r_{12})\vec r_{12} \times (\vec p_1 - \vec p_2) 
\varphi_s^*(\vec r_1) \varphi_p^{m' *}(\vec r_2 ) \varphi_p^{m} (\vec r_1) 
\varphi_s(\vec r_2)
\,.
\end{eqnarray}
Their scaling is ${\cal J}_i \sim O(N_c^0)$ since they depend 
only on the wave functions of the pair of interacting
quarks, but not on the total number of quarks.

The reduced matrix elements $\langle {\cal L}^s_S \rangle, 
\langle {\cal L}^s_{MS}\rangle$
scale like $O(N_c)$. Using this in Eq.~(\ref{Vsosymm}), one can see that the 
 contribution of the first term to the mass operator
 scales like $O(\alpha_s N_c^0) \sim O(N_c^{-1})$, while the
second  term gives a contribution of leading order 
$O(\alpha_s N_c) \sim O(N_c^{0})$ proportional to $\vec s_1 \cdot \vec L$, 
together with another power suppressed contribution.

Finally, we present also the contribution of the antisymmetric 
piece of the spin-orbit interaction to the mass operator. 
According to Eq.~(\ref{Oodd}), this is decomposed into
$MS\otimes  MS + A'\otimes A'$  under $S_N^{orb} \times
S_N^{sf}$.

We start by computing the $MS\otimes  MS$ term, which contributes
\begin{eqnarray}
\langle V_{so}^{anti} \rangle \equiv
\frac{\langle B | V_{so}^{anti} |B \rangle}{\langle B | B \rangle} =
\frac{N-2}{N^2} \langle \vec {\cal L}_{MS}^a \rangle 
\langle \vec {\cal O}_{MS}^a \rangle  \,.
\end{eqnarray}
The spin-flavor reduced matrix element can be obtained as explained in
Appendix~D, with the result
\begin{eqnarray}
\langle \vec {\cal O}_{MS}^a \rangle = \frac{N}{N-2} \langle \Phi(SI)|
N\vec s_1 - \vec S | \Phi(SI)\rangle \,,
\end{eqnarray}
and the orbital reduced matrix element is
\begin{eqnarray}
\langle \vec {\cal L}_{MS}^a \rangle = 
N {\cal J}_{\rm dir}^a (\vec L)_{m' m} \,.
\end{eqnarray}
The overlap integral is given explicitly by
\begin{eqnarray}
 (\vec L)_{m' m} {\cal J}^a_{\rm dir} &=& \int d\vec r_1 d\vec r_2
 \vec {\cal L}_{12}^a(r_{12}) 
 |\varphi_s(\vec r_1)|^2 \varphi_p^{m' *}(\vec r_2) \varphi_p^{m} (\vec r_2) \,.
\end{eqnarray}

This gives the total result for the $MS\times MS$ projection of the 
antisymmetric piece of the spin-orbit interaction
\begin{eqnarray}\label{VsoantiMS}
\langle V_{so}^{anti} \rangle = 
{\cal J}^a_{\rm dir} \Big(
N \vec s_1 \cdot \vec L - \vec S \cdot \vec L \Big) \,.
\end{eqnarray}

Finally, we consider also the $A'$ projection of the antisymmetric 
spin-orbit interaction. 
According to Eq.~(\ref{OA}), the $A'$ projection of the
spin-flavor operator ${\cal O}_{ij}^a = \vec s_i - \vec s_j$, vanishes
\begin{eqnarray}
{\cal O}_{jk}^{A'} = 2(\vec s_1 - \vec s_j) + 2(\vec s_j - \vec s_k) - 2(\vec s_1-\vec s_k)
= 0\,,
\end{eqnarray}
such that the $A'$ operators do not contribute.
In addition, for $N=3$ the orbital $A'$ contribution vanishes because of $T-$reversal
invariance. We give the detailed argument in the following, since it is independent
on the spin-flavor structure of the operator.

The $A'$ projection of $\vec {\cal L}_{ij}^a$ is given by
Eq.~(\ref{OA}) and can be taken as  $[\vec {\cal L}]^{A'} =
{\cal L}_{12}^a + {\cal L}_{23}^a - {\cal L}_{13}^a $. 
On the two-dimensional space of the $MS$ orbital states $\chi_i^m$, its 
matrix elements have the form
\begin{eqnarray}
\langle \chi_i^{m'} | [\vec {\cal L}_{ij}^a]^{A'} | \chi_j^m \rangle = \left(
\begin{array}{cc}
0 & -3{\cal J}_{\rm exc}^a  \\
3{\cal J}_{\rm exc}^a & 0 \\
\end{array}
\right)_{ij} (\vec L)_{m' m} \,,
\end{eqnarray}
where the exchange integral is defined as 
\begin{eqnarray}
(\vec L)_{m' m} {\cal J}^a_{\rm exc} = \int d\vec r_1 d\vec r_2
\vec {\cal L}_{12}^a(r_{12}) 
\varphi_s^*(\vec r_1) \varphi_p^{m' *}(\vec r_2 ) \varphi_p^{m} (\vec r_1) 
\varphi_s(\vec r_2)  \,.
\end{eqnarray}
Hermiticity of the $\vec {\cal L}_{12}^a$ operator implies that the overlap
integral is purely imaginary, $\mbox{Re }{\cal J}^a_{\rm exc}=0$.
However, $T-$reversal invariance of the Hamiltonian forbids imaginary
terms in the mass operator, such that the overlap integral must vanish.
For $N>3$, a second overlap integral can contribute, which can
be nonvanishing.

In summary, the mass operator of the orbitally excited baryons in the
nonrelativistic quark model with the gluon exchange potential 
Eq.~(\ref{Hamilton}) is given by 
\begin{eqnarray}\label{Mass0}
M = c_0 {\bf 1} - g_s^2 \frac{N_c+1}{2N_c} \Big(
a \vec S^2 + b \vec s_1 \cdot \vec S_c + c L_2^{ab} S^a S^b + 
d L_2^{ab} S_c^a  s_1^b + e \vec L\cdot \vec S  + f \vec L \cdot
\vec s_1 \Big) + \cdots \,.
\end{eqnarray}
The spin-flavor operators are understood to act on the CCGL-type state
$\Phi(SI)$, for which the excited quark is quark number 1. 
The ellipses
denote operators transforming in the $E$ irrep, which appear only for
$N > 3$, and are considered in the next section.

The coefficients $a-f$ are given by linear combinations of the orbital overlap
integrals introduced above
\begin{eqnarray}
a &=& \frac{1}{N(N-1)} \Big[\frac{(N-1)(N-2)}{2} {\cal I}_s + 
(N-1) {\cal I}_{\rm dir} - {\cal I}_{\rm exc}\Big]  \nonumber \\
& &\hspace{2cm} - \frac{1}{N(N-2)} 
  \Big[(N-2) ({\cal I}_{\rm dir} - {\cal I}_s) - 2 {\cal I}_{\rm exc}\Big] \,, 
   \\
b &=& \frac{1}{N-2} 
\Big[(N-2) ({\cal I}_{\rm dir} - {\cal I}_s) - 2 {\cal I}_{\rm exc}\Big]  \,, \\
c &=& \frac{2}{N(N-1)} \Big[(N-1) {\cal K}_{\rm dir} - {\cal K}_{\rm exc}\Big] - 
\frac{2}{N(N-2)} \Big[(N-2) {\cal K}_{\rm dir} - 2{\cal K}_{\rm exc}\Big] \,, \\
d &=& \frac{2}{N-2} \Big[(N-2) {\cal K}_{\rm dir} - 2 {\cal K}_{\rm exc}\Big] \,,\\
e &=& \frac{3}{N} 
\Big[(N-1){\cal J}_{\rm dir}^s-{\cal J}_{\rm exc}^s\Big] 
- \frac{3}{2 N} \Big[(N-2){\cal J}_{\rm dir}^s-2{\cal J}_{\rm exc}^s\Big] + 
    \frac{1}{2} {\cal J}_{\rm dir}^a \,, \\
f &=& \frac{3}{2}\Big[(N-2){\cal J}_{\rm dir}^s-2{\cal J}_{\rm exc}^s\Big]
- \frac{N}{2} {\cal J}_{\rm dir}^a  \,.
\end{eqnarray}
The matrix elements of the operators on quark model states can be found in the
Appendix of Ref.~\cite{Carlson:1998vx} for any values of $L$ and $N_c$.

The expression Eq.~(\ref{Mass0}) holds regardless of the orbital wave functions,
and summarizes in a compact operator form the most general result for the 
masses of the orbitally excited states with $MS$ spin-flavor symmetry, in the
presence of one-gluon exchange quark-quark interactions.
Although we used in deriving these results the Hartree approximation, they 
have a more general validity. The Hartree approximation is an useful device
for computing the leading $N_c$ dependence of the orbital reduced matrix elements,
but is not essential for the arguments leading to the spin-flavor structure
of the mass operator. Expressing the
latter directly in terms of the orbital reduced matrix elements will 
produce an expression similar to Eq.~(\ref{Mass0}), for which the $N$ 
dependence is implicit in the reduced matrix elements of the orbital operators.

\begin{table}
\begin{tabular}{|c|c|c|c||c|c|}\hline
 & \multicolumn{3}{|c||}{Symmetric} & \multicolumn{2}{|c|}{Antisymmetric}\\
 & \multicolumn{3}{|c||}{2-body} & \multicolumn{2}{|c|}{2-body}\\
\cline{2-6}
\hspace{1cm} & \,\, S\,\,  & \, MS\, & \,\,E\,\, & \, MS\, & A$^\prime$ \\\hline\hline
$V_{ss}$ & 1  & 1 & 1 & $-$ & $-$ \\
$V_{q}$ & 1  & 1 & 1 & $-$ & $-$ \\
$V_{so}$ & 1  & 1 & 0 & 1 & 0 \\
\hline
\end{tabular}
\caption{Summary of the spin-flavor contributions to the mass operator of the $L=1$ baryons
arising from each of the three terms in the interaction Hamiltonian (for both gluon-exchange
and Goldstone boson exchange interactions). The table counts
the independent structures, corresponding to each irrep of $S_N$. The dashes represent
terms forbidden by $S_N$, and the 0 show operators absent for other reasons. The $E$ operators
contribute only for $N_c > 3$. }
\label{table_redme}
\end{table}

\subsection{Goldstone boson exchange interaction}

We consider in this section the mass operator of the orbitally excited
baryons in a second model for the quark-quark interaction. 
In Ref.~\cite{Glozman:1995fu} it was suggested that
pion-exchange mediated quark-quark interactions can reproduce better
the observed mass spectrum of these states. The physical idea is that 
at the energy scales of quarks inside a hadron, the appropriate degrees
of freedom are quarks, gluons and the Goldstone bosons of the broken chiral
group $SU(2)_L \times SU(2)_R \to SU(2)$ \cite{Manohar:1983md}.
The exchange of Goldstone bosons changes the short distance form of the
quark-quark interactions, and introduces a different spin-flavor structure.

The interaction Hamiltonian of this model has the form
\begin{eqnarray}\label{HamiltonGR}
H = H_0 +  \frac{g_A^2}{f_\pi^2} \sum_{i<j}  \tilde V_{ij}
\end{eqnarray}
and has the same spin structure as the one-gluon interaction, but 
with additional explicit isospin dependence. $g_A$ is a quark-pion coupling
which scales like $O(N_c^0)$ with the number of colors $N_c$, and 
$f_\pi^2 \sim O(N_c)$.
The two-body interactions $\tilde V_{ij}$ include the following contributions:
spin-spin interaction $\tilde V_{ss}$, a tensor interaction 
$\tilde V_q$ and the spin-orbit interaction $\tilde V_{so}$. 
We write these interaction terms as 
\cite{Collins:1998ny}, 
\begin{eqnarray}
\tilde V_{ss} &=& \sum_{i < j=1}^N g_0(r_{ij}) \vec s_i \cdot \vec s_j t^a_i t^a_j 
\,, \\
\tilde V_{q} &=& \sum_{i < j=1}^N g_2(r_{ij}) 
\Big[3(\hat r_{ij} \cdot \vec s_i)(\hat r_{ij} \cdot \vec s_j)
 -  (\vec s_i \cdot \vec s_j) \Big] t^a_i t^a_j \,, \\
\tilde V_{so} &=& \sum_{i < j=1}^N g_1(r_{ij}) \Big[ 
(\vec r_{ij} \times \vec p_i ) \cdot \vec s_i - (\vec r_{ij} \times \vec p_j ) \cdot \vec s_j \\
& & \hspace{1.5cm}+ 2 (\vec r_{ij} \times \vec p_i ) \cdot \vec s_j - 2 
(\vec r_{ij} \times \vec p_j ) \cdot \vec s_i \Big] t^a_i t^a_j  \,, \nonumber
\end{eqnarray}
where the isospin generators are $t^a = \frac12 \tau^a$, and $g_i(r_{ij})$
are unspecified functions.

The mass operator of the orbitally excited baryons with this interaction
potential can be computed using the approach presented for the one-gluon
exchange interaction. The orbital matrix elements 
have the same form (although with different overlap integrals), while the
spin-flavor operators are different. 

We give in some detail the projection of the spin-spin interaction 
$\tilde V_{ij} = g_0(r_{ij}) \vec s_i \cdot \vec s_j t_i^a t_j^a$ onto 
operators transforming as irreps of $S_N^{sf}$. The symmetric projection
of the spin-flavor operator 
${\cal O}_{ij} = \vec s_i \cdot \vec s_j t_i^a t_j^a$ is
\begin{eqnarray}
{\cal O}_S = \sum_{i<j} {\cal O}_{ij} = \frac12 (G^{ka} G^{ka} - \frac{9}{16}N)
\,,
\end{eqnarray}
where we denoted $G^{ka} = \sum_{i=1}^N s_i^k t^a_i$. The $G^2$ operator
can be expressed in terms of the SU(4) Casimir using the identity (for
$F=2$ light quark flavors)
\begin{eqnarray}
G^{ka} G^{ka} = \frac{1}{16} N(3N+4) - \frac14 \vec S^2 - \frac14 T^a T^a
\,.
\end{eqnarray}

The reduced matrix element of the $MS$ projection can be found as explained
in Appendix~D, and is given by
\begin{eqnarray}
\langle {\cal O}_{MS}\rangle = \frac{1}{N-2} \Big(
N g_1^{ka} G_c^{ka} - G^{ka} G^{ka} + \frac{9}{16} N \Big) \,,
\end{eqnarray}
The $g_1 G_c$ terms can be reduced to simpler operators using the reduction
rule Eq.~(4.5) in Ref.~\cite{Carlson:1998vx}, which for $F=2$ gives
\begin{eqnarray}\label{reductionrule}
g_1^{ka} G_c^{ka} = - \frac{1}{16}(N+3) - \frac14 \vec s_1 \cdot \vec S_c
- \frac14 t_1^a T_c^a \,.
\end{eqnarray}
Finally, the operators have to expressed as sums of terms acting on the 
core and ``excited'' quark.

Collecting the contributions of all terms in the Hamiltonian 
we find the following general 
result for the mass operator in the Goldstone boson exchange model
\begin{eqnarray}\label{Mass1}
M &=& c_0 {\bf 1} +  \frac{g_A^2}{f_\pi^2} \Big(
a \vec S_c^2 + b \vec s_1 \cdot \vec S_c + c t_1^a T_c^a +
d L_2^{ij} g_1^{ia} G_c^{ja} + 
e L_2^{ij} \{S_c^i \,, S_c^j \} \\
& & \hspace{2cm} + f L^i S_c^i  + g  L^i t_1^a G_c^{ia} + h L^i g_1^{ia} T_c^a
 \Big) \nonumber \,.
\end{eqnarray}

The coefficients $a-h$ are given by linear combinations of the orbital overlap
integrals 
\begin{eqnarray}
a &=& - \frac{1}{2N(N-1)} \Big[ \frac12 (N-1)(N-2) {\cal I}_s +
(N-1) {\cal I}_{\rm dir} - {\cal I}_{\rm exc}\Big] \nonumber \\
& &\hspace{2cm} + \frac{1}{2N(N-2)} 
  \Big[(N-2) ({\cal I}_{\rm dir} - {\cal I}_s) - 2 {\cal I}_{\rm exc}\Big]\,, \\
b &=& c = - \frac{1}{2N(N-1)} \Big[ \frac12 (N-1)(N-2) {\cal I}_s +
(N-1) {\cal I}_{\rm dir} - {\cal I}_{\rm exc}\Big] \nonumber \\
& &\hspace{2cm} - \frac{1}{4N} 
  \Big[(N-2) ({\cal I}_{\rm dir} - {\cal I}_s) - 2 {\cal I}_{\rm exc}\Big]\,, \\
d &=& \frac{4}{N(N-1)} \Big[(N-1) {\cal K}_{\rm dir} - {\cal K}_{\rm exc}\Big] +
\frac{2}{N} \Big[(N-2) {\cal K}_{\rm dir} - 2{\cal K}_{\rm exc}\Big]\,, \\
e &=& \frac{1}{4N(N-1)} \Big[(N-1) {\cal K}_{\rm dir} - {\cal K}_{\rm exc}\Big] -
\frac{1}{4N(N-2)} \Big[(N-2) {\cal K}_{\rm dir} - 2{\cal K}_{\rm exc}\Big]\,, \\
f &=&  
\frac{3(N-2)}{4N(N-1)}\Big[(N-1){\cal J}_{\rm dir}^s-{\cal J}_{\rm exc}^s\Big]
-  \frac{3}{4N}
    \Big[(N-2){\cal J}_{\rm dir}^s-2{\cal J}_{\rm exc}^s\Big] \,, \\
g &=& \frac{3}{N(N-1)} \Big[(N-1){\cal J}_{\rm dir}^s-{\cal J}_{\rm exc}^s\Big]+  \frac{3}{2N}
\Big[(N-2){\cal J}_{\rm dir}^s-2{\cal J}_{\rm exc}^s\Big] 
   + \frac{1}{2} {\cal J}_{\rm dir}^a \,, \\
h &=& \frac{3}{N(N-1)} \Big[(N-1){\cal J}_{\rm dir}^s-{\cal J}_{\rm exc}^s\Big] 
+  \frac{3}{2N}
\Big[(N-2){\cal J}_{\rm dir}^s-2{\cal J}_{\rm exc}^s\Big] 
   - \frac{1}{2} {\cal J}_{\rm dir}^a \,.
\end{eqnarray}

Comparing with the mass operator formula for the one-gluon exchange
interaction Eq.~(\ref{Mass0}), we find that more effective operators are 
present for this case (8 vs. 6). Note that two of the coefficients 
are equal ($b=c$).  The total number of unknown constants
is the same in both cases and is given by the seven reduced matrix elements 
of the orbital operators. An additional simplification occurs for the
one-gluon interaction case, where the MS projections of the symmetric
and antisymmetric spin-orbit interaction give the same operator (compare
the second term in Eq.~(\ref{Vsosymm}) and Eq.~(\ref{VsoantiMS})). 
This reduces the number of independent spin-flavor structures for this case 
from seven to six.

\subsection{The $E$ operators}

In addition to the operators considered so far, transforming in the
$S, MS$ and $A'$ irreps, there are additional operators transforming in the $E$
irrep of $S_N$ which appear only for $N_c > 3$. They are introduced only by the 
spin-spin $V_{ss}$ and tensor $V_q$ interactions in the Hamiltonian. In this 
Section we study their spin-flavor structure.

We start by considering the spin-spin 
interaction ${\cal O}_{ij} =
\vec s_i \cdot \vec s_j$. The $E$ projection of this operator is given by
\begin{eqnarray}
{\cal O}_{abc}^E = (\vec s_1 - \vec s_b) \cdot (\vec s_a - \vec s_c) \,,
\end{eqnarray}
where $a,b,c$ are the integers appearing in the standard Young tableaux
in the order $[1c\cdots][ab]$. It turns out that it is impossible to
express the reduced matrix element of $E$ operators as
matrix elements on the CCGL-type state $\Phi(SI)$. 
The reason is that the diagonal matrix elements of the ${\cal O}^E_{abc}$ operators 
on the state $\Phi(SI)$ vanish. This can be seen by symmetrizing the operator 
${\cal O}^E_{abc}$ under permutations of the $N-1$ quarks different from quark 1, which
gives a vanishing result\footnote{In group theoretic language, this is due
to the fact that any $E$ operator transforms as $MS_{N-1} + E_{N-1}$ under the $S_{N-1}$
subgroup of $S_N$ which leaves quark 1 unchanged. Since $\Phi_1(SI)$ is a singlet
under this subgroup, the matrix elements of $E_{N}$ on this state must vanish.}.

On the other hand, the off-diagonal matrix elements of ${\cal O}_{abc}^E$ 
between $CCGL-$type states with different excited quarks, e.g. quarks 1 and 2,
are nonvanishing. We will denote them as $\Phi_1(SI)\equiv \Phi(SI)$ and 
$\Phi_2(SI)$; they are expressed in terms of the basis states $\phi_i$ as 
\begin{eqnarray}\label{Phi12}
|\Phi_1(SI)\rangle &=& \frac{1}{\sqrt{N(N-1)}}\Big(
\phi_2 + \phi_3 + \cdots + \phi_N\Big) \,,\\
|\Phi_2(SI)\rangle &=& \frac{1}{\sqrt{N(N-1)}}\Big(
-(N-1)\phi_2 + \phi_3 + \cdots + \phi_N
\Big)\nonumber
\end{eqnarray}

Consider for definiteness the matrix element of
${\cal O}^E_{234} = (\vec s_1 - \vec s_3)\cdot (\vec s_2 - \vec s_4)$, 
which can be written using the Wigner-Eckart theorem as
\begin{eqnarray}\label{CCGLE1}
\langle \Phi_1(SI) | {\cal O}_{234}^E |\Phi_2(SI) \rangle = \frac{N}{2(N-1)}
\langle {\cal O}_E \rangle
\end{eqnarray}
where the Clebsch coefficient has been computed using Eq.~(\ref{WEOE}). 
This shows that the
reduced matrix element of the $E$ operators $\langle {\cal O}_E \rangle$ 
can be expressed as the
off-diagonal matrix elements between $CCGL-$type states with different
excited quarks.

The operator on the left-hand side of Eq.~(\ref{CCGLE1})
can be put in a simpler form by noting
that both states $\Phi_{1,2}(SI)$ are symmetric under any permutation of the
`core' quarks $(3,4,\cdots,N)$. This core contains all $N-2$ quarks which
are different from quarks 1,2. Therefore the only nonvanishing contribution
to the matrix element in (\ref{CCGLE1}) will come from the component of
${\cal O}_{234}^E$ which is symmetric under the core quarks. This component is given as
a sum over all permutations $\Pi_c$ of the core quarks $(3,4,\cdots, N)$
\begin{eqnarray}
{\cal O}_{234}^E \to \frac{1}{(N-2)!} \Sigma_{\Pi_c} \Pi_c {\cal O}_{234}^E
\end{eqnarray}

The sum over permutations can be computed explicitly for the spin-spin
interaction with the result, for arbitrary $N$, 
\begin{eqnarray}
\frac{1}{(N-2)!} \Sigma_{\Pi_c} \Pi_c {\cal O}_{234}^E = 
(\vec s_1 \cdot \vec s_2) - \frac{1}{N-2}
(\vec s_1+\vec s_2)\cdot \vec S_{cc} + \frac{2}{(N-2)(N-3)} 
\sum_{i<j=3}^N \vec s_i \cdot \vec s_j
\end{eqnarray}
where we denoted $S_{cc} = s_3 + s_4 + \cdots + s_N$ the $(N-2)$-quarks core 
spin. The sum in the last term
runs over all quark pairs in the core.

Combining this with Eq.~(\ref{CCGLE1}) one finds the reduced matrix element 
of the $E$ operator as an off-diagonal matrix element on CCGL states with 
different excited quarks
\begin{eqnarray}
\frac{N}{2(N-1)} \langle {\cal O}_E \rangle  &=& 
\langle \Phi_1(SI) | {\cal O}_{234}^E |\Phi_2(SI) \rangle \\
& &\hspace{-3cm} =
\langle \Phi_1(SI) |(\vec s_1 \cdot \vec s_2) - \frac{1}{N-2}
(\vec s_1+\vec s_2)\cdot \vec S_{cc} + 
\frac{1}{(N-2)(N-3)} (\vec S_{cc}^2  - \frac34 (N-2) )
|\Phi_2(SI) \rangle \nonumber
\end{eqnarray}
This specifies the reduced matrix of the operator $\langle {\cal O}_E\rangle$ as an
off-diagonal matrix element on CCGL-type states. 

The contribution of the $E$ component of the tensor interaction is computed 
in an analogous way. Combining everything we find the total contribution of
the $E$ operators to the mass operator Eq.~(\ref{Mass0}) in the
one-gluon exchange model as $-g_s^2\frac{N_c+1}{2N_c}\delta M_E$ with 
\begin{eqnarray}\label{ME0}
\delta M_E &=& g \Big( (\vec s_1 \cdot \vec s_2) - \frac{1}{N-2}
(\vec s_1+\vec s_2)\cdot \vec S_{cc} + 
\frac{1}{(N-2)(N-3)}[S_{cc}^2 - \frac34(N-2)] \Big) \\
&+& h L_2^{ab}\Big( s_1^a s_2^b - \frac{1}{N-2}
(s_1^a+s_2^a) S_{cc}^b + 
\frac{1}{2(N-2)(N-3)} \{ S_{cc}^a \,, S_{cc}^b\} \Big)\nonumber
\end{eqnarray}
with
\begin{eqnarray}
g = (N-3){\cal I}_{\rm exc}\,,\qquad
h = (N-3){\cal K}_{\rm exc}
\end{eqnarray}
The matrix element of the operators in (\ref{ME0}) is understood to be
taken between the off-diagonal CCGL-type spin-flavor 
states $\langle \Phi_1(SI)|$ and
$|\Phi_2(SI)\rangle$ states defined as in Eq.~(\ref{Phi12}).

For the purposes of $N$ power counting, it is convenient to express the
operators in Eq.~(\ref{ME0}) as diagonal matrix elements on the $\Phi_1(SI)$
state. This can be done using the relation
\begin{eqnarray}
|\Phi_2(SI)\rangle = P_{12} |\Phi_1(SI)\rangle
\end{eqnarray}
together with the explicit representation of the 
transposition operator $P_{12}$ on a system of spin-isospin $1/2$ quarks
\begin{eqnarray}
P_{12} = \frac14 (1 + 4\vec s_1 \cdot \vec s_2) (1 + 4 t_1^a t_2^a)\,.
\end{eqnarray}

We illustrate the method of calculation by giving the details of the
off-diagonal matrix element of the first term in Eq.~(\ref{ME0}). This can
be written as
\begin{eqnarray}
\langle \Phi_1(SI) | \vec s_1 \cdot \vec s_2 |\Phi_2(SI)\rangle &=& 
\langle \Phi_1(SI) | \vec s_1 \cdot \vec s_2 P_{12} |\Phi_1(SI)\rangle\\
&=& 
\frac{1}{N-1}\langle \Phi_1(SI) | 
\sum_{k=2}^N\vec s_1 \cdot \vec s_k P_{1k} |\Phi_1(SI)\rangle \nonumber
\end{eqnarray}
In the last step we used the symmetry of the wave function $\Phi_1(SI)$ under
the $S_{N-1}^{sp-fl}$ subgroup of $S_N^{sp-fl}$ acting only on
the quarks $2,3,\cdots, N$, and projected the operator 
$\vec s_1 \cdot \vec s_2 P_{12}$ onto its symmetric component under this 
subgroup. The sum over $k$ can be performed in closed form with the result
\begin{eqnarray}
\langle \Phi_1(SI) | \vec s_1 \cdot \vec s_2 |\Phi_2(SI)\rangle &=&
\frac{1}{4(N-1)}\langle \Phi_1(SI) | 
- \vec s_1 \cdot \vec S_c + \frac34 (N-1) - 4 g_1^{ia} G_c^{ia} + 
3 t_1^a T_c^a |\Phi_1(SI)\rangle \nonumber\\
&=& \langle \Phi_1(SI) | \frac{1}{N-1} t_1^a T_c^a + \frac{N}{4(N-1)} | \Phi_1(SI)
\rangle\,,
\end{eqnarray}
where we used the reduction rule Eq.~(\ref{reductionrule}) in the last step to
eliminate $\vec s_1 \cdot \vec S_c, g_1^{ia} G_c^{ia}$ in favor of
$t_1^a T_c^a$.

The remaining terms in Eq.~(\ref{ME0}) can be computed in a similar way. After a lengthy
calculation one finds for the off-diagonal matrix element of the first term,
arising from the spin-spin interaction,
\begin{eqnarray}\label{E-CCGL}
& &\langle \Phi_1(SI) |(\vec s_1 \cdot \vec s_2) - \frac{1}{N-2}
(\vec s_1+\vec s_2)\cdot \vec S_{cc} + 
\frac{1}{(N-2)(N-3)} (\vec S_{cc}^2  - \frac34 (N-2) )
|\Phi_2(SI) \rangle = \nonumber\\
& & \langle \Phi_1(SI) |\frac{3N-13}{4(N-2)(N-3)} t_1 T_c - 
\frac{N^2+4N+11}{4(N-1)(N-2)(N-3)} s_1 S_c -  \frac{2}{(N-2)(N-3)} g_1 G_c\nonumber \\
& & \hspace{0.5cm}- \frac{N+3}{4(N-1)(N-2)(N-3)} S_c^2 - \frac{1}{(N-2)(N-3)} g_1S_c T_c
+ \frac{N}{4(N-2)} |\Phi_1(SI) \rangle \,.
\end{eqnarray}
In writing the final result we assumed two light flavors $F=2$, which 
allows the use
of the identity $S_c^i G_c^{ia} = \frac14(N+1) T_c^a$. 

The matrix element of the second term in  Eq.~(\ref{ME0}), representing the
contribution of the $E$ projection of the quadrupole interaction $V_{q}$, can be
expressed as a diagonal matrix element on $\Phi_1(SI)$ in a similar way, with the result
\begin{eqnarray}\label{E-CCGL2}
& &L_2^{ij} \langle \Phi_1(SI) |\{s_1^i, s_2^j\} - \frac{1}{N-2}
\{(s_1^i + s_2^i), S_{cc}^j\} + 
\frac{1}{(N-2)(N-3)} \{S_{cc}^i , S_{cc}^j \}
|\Phi_2(SI) \rangle = \nonumber\\
& & \hspace{0.5cm} L_2^{ij}\langle\Phi_1(SI)| \frac{2N-5}{(N-2)(N-3)} \{g_1^{ia}, G_c^{ja}\} +
\frac{N+1}{4(N-1)(N-2)} \{s_1^i, S_c^j\} \nonumber \\
& &\hspace{1.5cm} -  \frac{N+3}{4(N-1)(N-2)(N-3)} \{S_c^i, S_c^j\}
 - \frac{1}{(N-2)(N-3)} t_1^a \{ S_c^i, G_c^{ja}\} \\
& &\hspace{1.5cm}
- \frac{1}{(N-2)(N-3)} \{ g_1^{ia},S_c^j\} T_c^a 
|\Phi_1(SI) \rangle 
\nonumber\,.
\end{eqnarray}

A few comments are in order about the form of these results. Note that the
$E$ operators arising from the spin-spin and tensor interactions
introduce a nontrivial flavor dependence (manifested through the operators 
$t_1 T_c, g_1 G_c$) which is not observed in the $S,MS$
projections of these interactions. Furthermore, in addition to the operators 
encountered so far, 
three new structures are introduced by the $E$ operators, given by 
$g_1^{ia} S_c^i T_c^a$ and $L_2^{ij} t_1^a \{ S_c^i, G_c^{ja}\}$,
$L_2^{ij} \{g_1^{ia}, S_c^j\} T_c^{a}$.
The operator $L_2^{ij}\{g_1^{ia} , G_c^{ja}\}$ contributes at $O(N_c^0)$, 
and the remaining ones start at order $O(N_c^{-1})$.

\section{Matching onto the $1/N_c$ expansion}

The spin-flavor structure of the matrix elements of the one-gluon interaction 
Eq.~(\ref{Mass0})
matches a subset of the operators appearing in the $1/N_c$ expansion of the 
mass operator 
of the orbitally excited states. We adopt here the
basis of operators of Ref.~\cite{Carlson:1998vx}. Working to order $1/N_c$, the
most general set of operators for the mass of these states is
\begin{eqnarray}
\hat M = c_1 N_c {\bf 1} + c_2 L^i s^i + 
c_3 \frac{3}{N_c} L_2^{ij} g^{ia} G_c^{ja}
+ \sum_{i=4}^8 c_i {\cal O}_i \,.
\end{eqnarray} 
The terms proportional to $c_{2,3}$ contribute at order $O(N_c^0)$, and the remaining
operators proportional to $c_{4-8}$ are of order $1/N_c$. A
complete basis of subleading operators
can be chosen as \cite{Carlson:1998vx}
\begin{eqnarray}
{\cal O}_4 &=& L^i s^i + \frac{4}{N_c+1} L^i t^a G_c^{ia} \,,\qquad
{\cal O}_5 = \frac{1}{N_c} L^i S_c^{i} \,,\qquad
{\cal O}_6 = \frac{1}{N_c} S_c^2 \,,\\
{\cal O}_7 &=& \frac{1}{N_c} s^i S_c^i \,,\qquad
{\cal O}_8 = \frac{1}{N_c} L_2^{ij} \{s^i\,, S_c^j\}\,,\qquad
{\cal O}_9 = \frac{1}{N_c} L^i g^{ia} T_c^a \,, \nonumber\\
{\cal O}_{10} &=& \frac{1}{N_c} t^a T_c^a \,,\qquad
{\cal O}_{11} = \frac{1}{N_c^2} L_2^{ij} t^a \{S_c^i\,, G_c^{ja}\} \,. \nonumber
\end{eqnarray}

Matching the one gluon exchange quark-quark interaction we find a
leading order $O(N_c^0)$ contribution to the mass coming from the 
spin-orbit interaction
\begin{eqnarray}\label{OG-prediction}
c_2 &=& -\frac{g_s^2 N_c}{4} \Big( 3 {\cal J}_{\rm dir}^s - 
 {\cal J}_{\rm dir}^a \Big) \,,\qquad
c_3 = 0\,.
\end{eqnarray}
This confirms in a direct way the prediction obtained from the $1/N_c$ expansion
of the breaking of the SU(4) spin-flavor symmetry at leading order in $N_c$
\cite{Goity:1996hk,Carlson:1998vx}.
The nonrelativistic quark model with gluon mediated quark interactions displays
the same breaking phenomenon.

The coefficients of the $O(N_c^{-1})$ operators are given by
\begin{eqnarray}
& &c_4 = 0 \,, \qquad\qquad
c_5 = - \frac{g_s^2 N_c}{4} \Big( 3 {\cal J}_{\rm dir}^s  + {\cal J}_{\rm dir}^a \Big)
\,, \\
& &c_6 = -\frac{g_s^2 N_c}{4} {\cal I}_s \,,\qquad 
c_7 = -\frac{g_s^2 N_c}{2} {\cal I}_{\rm dir}  \,,\qquad
c_8 = -\frac{g_s^2 N_c}{2}  {\cal K}_{\rm dir} \nonumber
\end{eqnarray}
and $c_{9,10,11} = 0 $. 
In addition, the $E$ operators contribute to the coefficients $c_{6,7,10}$,
as shown in Eq.~(\ref{E-CCGL}), and to $c_{3,8,11}$ as seen in 
Eq.~(\ref{E-CCGL2}).
They also introduce two operators not present for $N=3$, 
$g_1^{ia} S_c^i T_c^a$ and $L_2^{ij} \{g_1^{ia}, S_c^j\} T_c^{a}$, 
contributing at order $O(N_c^{-2})$,
in agreement with Ref.~\cite{Carlson:1998vx}.

Several general comments can be made about these results.

The explicit calculation confirms the $N_c$ power counting rules
given in Ref.~\cite{Goity:1996hk}, giving the leading contribution of
each operator. The two-body quark interactions considered
here produce one-, two- and three-body (${\cal O}_{17} = \frac{1}{N_c^2}L_2^{ij}
\{ S_c^i\,, S_c^j \}$ of Ref.~\cite{Carlson:1998vx}, which correctly 
appears at order $O(1/N_c^2)$ in Eqs.~(\ref{Mass0}), (\ref{E-CCGL2}) ) effective 
operators in the $1/N_c$ expansion.
Higher order iterations of the interaction Hamiltonian will also generate more
higher order operators. 

An important conclusion following from this calculation is that 
operators with nontrivial permutation symmetry are indeed required by a correct 
implementation of the $1/N_c$ expansion. 
This disagrees with the $1/N_c$ expansion recently proposed by Matagne and 
Stancu \cite{Matagne:2006dj}, which does not allow for such operators.

A distinctive prediction of the one-gluon exchange potential is the vanishing
of the coefficient $c_3$ of one of the leading order operators 
 (up to contributions
from the $E$ operators, which appear only for $N>3$).
Expressed in terms of the
tower masses $M_0, M_1, M_2$, (see Ref.\cite{PiSc}) this relation is equivalent 
to a mass relation at 
leading order in $1/N_c$
\begin{eqnarray}
M_0 - \frac32 M_1 + \frac12 M_2 = O(N_c^{-1}) \ . 
\end{eqnarray}

We consider next the matching of the mass operator Eq.~(\ref{Mass1}) in the
model with quark-quark interactions
mediated by Goldstone boson exchange. For this case the coefficients of the
leading operators are 
\begin{eqnarray}
c_2 =-  \frac18 \tilde g_A^2 ( 3 {\cal J}_{\rm dir}^s + {\cal J}_{\rm dir}^a ) 
\,,\qquad
c_3  =  \frac23 \tilde g_A^2 {\cal K}_{\rm dir}
\end{eqnarray}
where we defined $\tilde g_A^2 = N_c g_A^2/f_\pi^2$ a coefficient
of order $ \sim O(N_c^0)$.

The coefficients of the subleading $O(N_c^{-1})$ operators are
\begin{eqnarray}
c_4 &=& \frac18 \tilde g_A^2 ( 3 {\cal J}_{\rm dir}^s + {\cal J}_{\rm dir}^a ) 
\,,\qquad
c_5=0 \,,\qquad
c_6 = -\frac14 \tilde g_A^2  {\cal I}_{\rm s}\,,\\
c_7 &=& c_{10}= -\frac14 \tilde g_A^2 {\cal I}_{\rm dir} \,,\qquad
c_8 = c_{11} = 0 \,,\qquad
c_{9} = \frac12 \tilde g_A^2 ( 3 {\cal J}_{\rm dir}^s - {\cal J}_{\rm dir}^a ) 
\,. \nonumber
\end{eqnarray}

For this case the coefficients of both leading operators can be of 
natural size.
There are predictions for the vanishing of the coefficients
of some subleading operators, and also one
relation between the coefficients of the leading and subleading operators
$c_2= -c_4 $. The Goldstone boson interaction also generates a 
three-body operator, ${\cal O}_{17} $ of
Ref.~\cite{Carlson:1998vx}, which correctly appears at 
order $O(1/N_c^2)$ in Eq.~(\ref{Mass1}).

The coefficients of the leading order operators $c_{1,2,3}$ have been
determined in  \cite{PiSc} from a fit to the masses of the nonstrange 
$L=1$ baryons, working at leading order in $1/N_c$. 
The results depend on the assignment of the observed baryons into the irreps
of the contracted symmetry (towers). There are four possible assignments, but
only two of them are favored by data. These two assignments give the
coefficients
\begin{eqnarray}
\mbox{assignment 1: } \qquad &&
c_2^{(0)} = 83\pm 14 \mbox{ MeV}\,,\qquad c_3^{(0)} = -188\pm 28 \mbox{ MeV} \\
\mbox{assignment 3: } \qquad &&
c_2^{(0)} = -12\pm 16 \mbox{ MeV}\,,\qquad c_3^{(0)} = 142\pm 38 \mbox{ MeV} 
\end{eqnarray}
Comparing with the predictions Eq.~(\ref{OG-prediction}) of the one-gluon 
quark-quark potential model, we see that there is no evidence in the data for
a suppression of the coefficient $c_3$ relative to $c_2$. For both assignments,
the coefficient $c_3$ is sizeable, a situation which favors the
Goldstone boson exchange model, or at least indicates that 
some kind of flavor dependent effective interactions cannot be neglected. 
A more detailed analysis, including also the
predictions for the subleading coefficients, will be presented elsewhere.

\section{Conclusions}

In this paper we analyzed the spin-flavor structure of excited
baryons containing $N_c$ quarks, from the perspective of the permutation 
group $S_N$ of $N=N_c$ objects. The group $S_N$ is the diagonal subgroup
of the product $S_N^{orb} \times S_N^{sf}$ of permutations
acting separately on the  orbital and spin-flavor wave functions respectively.

The permutation group imposes restrictive constraints on the form of the 
allowed spin-flavor operators contributing to any physical quantity, such as
masses and couplings. In this paper we discussed in detail the mass operator of
the orbitally excited baryons with $MS$ spin-flavor symmetry. 

In the quark model with a quark-quark interaction Hamiltonian $V$, 
the hadronic mass operator contains only those spin-flavor
operators which appear in the decomposition of the Hamiltonian under
$S_N \to S_N^{orb} \times S_N^{sf}$. This decomposition has the
generic form $V = \sum_R V^{orb}_R V^{sf}_R$, with $R$ 
irreps of the $S_N$ group, which depend on $V$. Taking the
matrix elements of the orbital operators $V^{orb}_R$ on hadronic
states replaces them with
reduced matrix elements, parameterizing in a general way our ignorance
about the orbital wave functions. The remaining dependence on the spin-flavor
degrees of freedom is displayed explicitly in operator form, regardless
of the unknown orbital wave functions. 

This approach is similar to the method used for
$N_c=3$ in Ref.~\cite{Collins:1998ny} to obtain the predictions
of the nonrelativistic quark model in a form independent of the orbital 
wave functions.
In addition to extending this result to arbitrary $N_c$, we focus
here on the transformation of the operators under the permutation
group $S_N^{sf}$ acting on the spin-flavor degrees of freedom,
which is crucial for the connection of these results with the $1/N_c$ expansion.

We constructed explicitly the $S_N \to S_N^{orb} \times S_N^{sf}$
decomposition of the most general two-body operator $V$, considering in 
particular the case of the hyperfine gluon-exchange and Goldstone boson 
exchange mediated
quark-quark interactions. The main results of this analysis are Eqs.~(\ref{Mass0}) 
and (\ref{Mass1}), which 
summarize in a compact operator form the most general structure of the
mass operator for the excited baryons allowed by the considered $q-q$
interactions.

The results for the mass operator match precisely the effective operators appearing
in the $1/N_c$ expansion for these states \cite{Goity:1996hk,Carlson:1998vx},
We find that the decomposition into
core and excited quark operators used previously in the literature on the
subject is both necessary, and a consequence of the $S_N$ symmetry.
We confirm by explicit calculation the $N_c$ counting rules of
Refs.~\cite{Goity:1996hk,Carlson:1998vx}. 
In particular we confirm that the effective spin-orbit 
interaction is of leading order
$O(N_c^0)$, as was correctly stated in Ref.~\cite{Goity:1996hk}.

Comparing with existing fits to the masses of nonstrange $L=1$ excited baryons,
we find that flavor dependent interactions cannot be neglected, and
may be necessary to supplement the gluon exchange model. This is 
in line with the chiral quark picture proposed in 
Ref.~\cite{Manohar:1983md}. The approach discussed here allows further
tests of the nature of the $q-q$ forces, as manifested through the hadronic
properties of these states.
We hope to report progress on this issue in the near future.

\section*{Acknowledgements}
 C.S. acknowledges financial support from
 Fundaci\'on S\'eneca and the hospitality of the Physics Department 
of the University of Murcia, Spain, where part of this work was done.

\newpage

\appendix
\section{The permutation group}

In this Appendix we give a few details about the permutation group of $N$ objects $S_N$, 
and its irreducible representations which are used in the main text.

The irreps of $S_N$ are identified by a partition $[n_1, n_2, \cdots , n_j]$ 
with $\sum_{i=1}^j n_i = N$. Each partition  can be represented as a Young 
diagram with $n_1$ boxes on the first row, $n_2$ on the second row, etc.

The irreps of $S_N$ which are relevant for the present work are given
in Table \ref{table1}. We list also the characters for the conjugacy class 
$1^\alpha 2^\beta 3^\gamma\cdots$ containing $\alpha$
1-cycles, $\beta$ 2-cycles, etc. Using these expressions and the 
orthogonality theorem for characters, the 
projection of any operator onto these irreps of $S_N$ can be computed 
explicitly.

\begin{table}[h]
\begin{eqnarray}\nonumber
\begin{array}{|c|c|c|c|}
\hline
\mbox{Name} & \mbox{Partition} & \mbox{Dim} & \mbox{Character}  \ 
\chi^{(R)}_{1^\alpha, 2^\beta, 3^\gamma,\cdots}\\
\hline
\hline
S & [N] & 1  & 1\\
MS & [N-1,1] & N-1  & \alpha-1 \\
E & [N-2,2] & \frac12N(N-3)  & \frac12 (\alpha - 1)(\alpha - 2) + \beta - 1 \\
A' & [N-2,1,1] & \frac12 (N-1)(N-2)  & \frac12 (\alpha - 1)(\alpha - 2) - \beta \\
\hline
\end{array}
\end{eqnarray}
\caption{Irreducible representations of $S_N$ contained in $MS \times MS$. } \label{table1}
\end{table}

The choice of a basis for a given irrep is not unique and several possible
choices are adopted in the literature. The most elegant choice is the 
orthogonal Yamanouchi-Katani
basis, which is constructed recursively in terms of the chain $S_{N-1} \subset
S_{N} \subset S_{N+1} \cdots$. However, this basis is not convenient for 
arbitrary $N$, since the matrices of the irreducible representations of $S_N$ 
can not be given
in a simple closed form.

In this paper we use the Young operator 
basis \cite{Hamm}. Consider a particular standard Young tableaux (SYT)
of the shape corresponding to a given irrep of $S_N$. This is a Young tableaux
with the numbers 1, 2, $\cdots $, N inserted such that the numbers increase
from the left to the right in each row, and increase as one moves down in each
column. We will denote a SYT by giving the sequence of indices in each row, e.g.
$[125][3][4]$ for a Young diagram with 3 rows in the $A'$ of $S_5$. 
Denote with $P$ the sum of all
horizontal permutations, which exchange only elements in the same row,
and with $Q$ the sum of all vertical permutations, which interchange 
only elements in the same column, multiplied with the parity of each permutation.

The Young operator $Y$ corresponding to a given Young tableau is defined as
\begin{eqnarray}
Y = QP \,.
\end{eqnarray} 
This can be applied to any state or operator to project out the component
with the desired transformation under $S_N$.

As an example we give the explicit construction of the basis of symmetric 
two-body operators ${\cal O}_{ij}$ transforming under irreps of $S_N$, for $N=5$. 
As shown in Sec.~\ref{Sec:NRQM}, 
these operators are decomposed as $\{ {\cal O}_{ij}\} = S + MS + E$ under 
the permutation group $S_N$.

The symmetric component is given by the sum over all components 
\begin{eqnarray}
{\cal O}^S = {\cal O}_{12} + {\cal O}_{13} + {\cal O}_{14} + {\cal O}_{15} +
{\cal O}_{23} + {\cal O}_{24} + {\cal O}_{25} + {\cal O}_{34} +
{\cal O}_{35} + {\cal O}_{45}  \,.
\end{eqnarray}

The $MS$ operators are constructed by applying the Young projection operators
to a certain operator $O_{ij}^{(0)}$. This is chosen such that the resulting
basis of operators satisfies the relations Eqs.~(\ref{phitransp}). 
Considering the Young tableaux $[1ab\cdots ][k]$
corresponding to $O_k^{MS}$, the starting operator must be chosen as
$O_{ij}^{(0)}=O_{1a}$. 
Thus, for $O_2^{MS}$ the Young projection operator is applied to $O_{13}$, and
for the remaining $O_{k\geq 3}^{MS}$, to $O_{12}$. As an illustration,
we give below the explicit $MS$ operator basis for $N=5$, where we show
also the standard Young diagram corresponding to each projector
\begin{eqnarray}
{\cal O}^{MS}_2 &=& {\cal O}_{13} + {\cal O}_{14} + {\cal O}_{15}
-{\cal O}_{23} - {\cal O}_{24} - {\cal O}_{25} 
 \,, \qquad [1345][2] \\
{\cal O}^{MS}_3 &=& {\cal O}_{12} + {\cal O}_{14} + {\cal O}_{15} -
 {\cal O}_{23} - {\cal O}_{34} - {\cal O}_{35}\,, \qquad [1245][3] \\
{\cal O}^{MS}_4 &=& {\cal O}_{12} + {\cal O}_{13} + {\cal O}_{15} -
 {\cal O}_{24} - {\cal O}_{34} - {\cal O}_{45}\,, \qquad [1235][4] \\
{\cal O}^{MS}_5 &=& {\cal O}_{12} + {\cal O}_{13} + {\cal O}_{14} -
 {\cal O}_{25} - {\cal O}_{35} - {\cal O}_{45}\,, \qquad [1234][5]  \,.
\end{eqnarray}

The $E$ projection contains five operators. They can be obtained
by acting with the Young projectors $Y(E)$ of the standard Young diagram
$[1a\cdots][bc]$ onto $O_{1a}$, and are given by
\begin{eqnarray}
{\cal O}^{E}_1 &=& {\cal O}_{12} - {\cal O}_{15} - {\cal O}_{24} +
{\cal O}_{45} \,, \qquad [123][45]\\
{\cal O}^{E}_2 &=& {\cal O}_{12} - {\cal O}_{15} - {\cal O}_{32} +
{\cal O}_{35} \,, \qquad [124][35]\\
{\cal O}^{E}_3 &=& {\cal O}_{12} - {\cal O}_{14} - {\cal O}_{23} +
{\cal O}_{34} \,, \qquad [125][34]\\
{\cal O}^{E}_4 &=& {\cal O}_{13} -
{\cal O}_{15} - {\cal O}_{23} + {\cal O}_{25}  \,, \qquad [134][25]\\
{\cal O}^{E}_5 &=& {\cal O}_{13} -
{\cal O}_{14} - {\cal O}_{23} + {\cal O}_{24}  \,, \qquad [135][24] \,.
\end{eqnarray}

These results can be used to express any interaction of $N=5$
particles interacting through symmetric two-body potentials which are factorized
into spin-isospin and orbital operators. Expressed
in terms of the operators transforming in the irreps of $S_N$, 
$\tilde O_k^T = (O_S, O_{2}^{MS},  O_{3}^{MS},  O_{4}^{MS},  O_{5}^{MS}, 
 O_{1}^{E},  O_{2}^{E},  O_{3}^{E},  O_{4}^{E},  O_{5}^{E})$, the
interaction Hamiltonian takes a block diagonal form 
\begin{eqnarray}
\sum_{i<j} {\cal O}_{ij} {\cal R}_{ij} = 
\sum_{k,l} \tilde {\cal O}_k  
\left(
\begin{array}{cccccccccc}
\frac{1}{10} & 0 & 0 & 0 & 0 & 0 & 0 & 0 & 0 & 0 \\
0 & \frac{4}{15} & - \frac{1}{15} & - \frac{1}{15} & - \frac{1}{15} & 
        0 & 0 & 0 & 0 & 0 \\
0 & - \frac{1}{15} & \frac{4}{15} & - \frac{1}{15} & - \frac{1}{15} &
        0 & 0 & 0 & 0 & 0 \\
0 & - \frac{1}{15} & - \frac{1}{15} & \frac{4}{15} & - \frac{1}{15} &
        0 & 0 & 0 & 0 & 0 \\
0 & - \frac{1}{15} & - \frac{1}{15} & - \frac{1}{15} & \frac{4}{15} & 
        0 & 0 & 0 & 0 & 0 \\
0 & 0 & 0 & 0 & 0 &
\frac12 & -\frac16 & -\frac16 & -\frac16 & \frac13 \\
0 & 0 & 0 & 0 & 0 &
-\frac16 & \frac12 & -\frac16 & -\frac16 & 0 \\
0 & 0 & 0 & 0 & 0 &
-\frac16 & -\frac16 & \frac12 & \frac16 & -\frac13 \\
0 & 0 & 0 & 0 & 0 &
-\frac16 & -\frac16 & \frac16 & \frac12 & -\frac13 \\
0 & 0 & 0 & 0 & 0 &
\frac13 & 0 & -\frac13 & -\frac13 & \frac23 \\
\end{array}
\right)_{kl}\tilde {\cal R}_l \,.
\end{eqnarray}
This is a particular case for $N=5$ of the general relation 
Eq.~(\ref{Oeven}) which is valid for any $N$.

We also present the construction of the operator basis for the
antisymmetric two-body operators. As mentioned, these operators
transform like the $MS + A'$ irreps.

The MS operators are given by
\begin{eqnarray}
O^{MS}_2 &=& 2O_{12} - O_{23} - O_{24} - O_{25} + O_{13} + O_{14} + O_{15}
\,,\qquad [1345][2] \\
O^{MS}_3 &=& 2O_{13} + O_{23} - O_{34} - O_{35} + O_{12} + O_{14} + O_{15}
\,,\qquad [1245][3]\\
O^{MS}_4 &=& 2O_{14} + O_{24} + O_{34} - O_{45} + O_{12} + O_{13} + O_{15}
\,,\qquad [1235][4]\\
O^{MS}_5 &=& 2O_{15} + O_{25} + O_{35} + O_{45} + O_{12} + O_{13} + O_{14}
\,,\qquad [1234][5] \,.
\end{eqnarray}

The $A'$ operators depend only on the indices in the first column of the SYT.
Considering the SYT $[1ab][c][d]$, the Young projector $Y(A')$ reduces
to the antisymmetrizer in the indices $[1cd]$, $A_{1cd} =
1-P_{1c}-P_{1d}-P_{cd} + P_{cd1} + P_{c1d}$, when applied to $O_{1c}$.
The basis of operators is
\begin{eqnarray}
O_1^{A'} &=& 2{\cal O}_{12} - 2{\cal O}_{13} + 2{\cal O}_{23} \,, \qquad [145][2][3]\\
O_2^{A'} &=& 2{\cal O}_{12} - 2{\cal O}_{14} + 2{\cal O}_{24} \,, \qquad [135][2][4]\\
O_3^{A'} &=& 2{\cal O}_{12} - 2{\cal O}_{15} + 2{\cal O}_{25} \,, \qquad [134][2][5]\\
O_4^{A'} &=& 2{\cal O}_{13} - 2{\cal O}_{14} + 2{\cal O}_{34} \,, \qquad [125][3][4]\\
O_5^{A'} &=& 2{\cal O}_{13} - 2{\cal O}_{15} + 2{\cal O}_{35} \,, \qquad [124][3][5]\\
O_6^{A'} &=& 2{\cal O}_{14} - 2{\cal O}_{15} + 2{\cal O}_{45} \,, \qquad [123][4][5]
\,.
\end{eqnarray}

The decomposition of a symmetric Hamiltonian containing antisymmetric two-body operators, 
expressed in the basis
$\tilde O_k^T = (O_{2}^{MS},  O_{3}^{MS},  O_{4}^{MS},  O_{5}^{MS}, 
 O_{1}^{A'},  O_{2}^{A'},  O_{3}^{A'},  O_{4}^{A'},  O_{5}^{A'}, O_6^{A'})$, has again
a block diagonal form 
\begin{eqnarray}
\sum_{i<j} {\cal O}_{ij} {\cal R}_{ij} = 
\sum_{k,l} \tilde {\cal O}_k  
\left(
\begin{array}{cccccccccc}
\frac{4}{25} & -\frac{1}{25} & -\frac{1}{25} & -\frac{1}{25} & 0 & 0 & 0 & 0 & 0 & 0 \\
-\frac{1}{25} & \frac{4}{25} & - \frac{1}{25} & - \frac{1}{25} & 0 & 
        0 & 0 & 0 & 0 & 0 \\
-\frac{1}{25} & - \frac{1}{25} & \frac{4}{25} & -\frac{1}{25} & 0 &
        0 & 0 & 0 & 0 & 0 \\
-\frac{1}{25} & - \frac{1}{25} & - \frac{1}{25} &  \frac{4}{25} & 0 & 
        0 & 0 & 0 & 0 & 0 \\
0 & 0 & 0 & 0 &  
\frac{3}{20} & -\frac{1}{20} &  -\frac{1}{20} &  \frac{1}{20} & \frac{1}{20} & 0 \\
0 & 0 & 0 & 0 &  
-\frac{1}{20} & \frac{3}{20} &  -\frac{1}{20} &  -\frac{1}{20} & 0 & \frac{1}{20} \\
0 & 0 & 0 & 0 &  
-\frac{1}{20} & -\frac{1}{20} &  \frac{3}{20} & 0 & -\frac{1}{20}  & -\frac{1}{20} \\
0 & 0 & 0 & 0 &  
\frac{1}{20} & -\frac{1}{20} & 0 &  \frac{3}{20} & -\frac{1}{20} & \frac{1}{20} \\
0 & 0 & 0 & 0 &  
\frac{1}{20} & 0 &  -\frac{1}{20} &  -\frac{1}{20} & \frac{3}{20} & -\frac{1}{20} \\
0 & 0 & 0 & 0 &  
0 & \frac{1}{20} &  -\frac{1}{20} &  \frac{1}{20} & -\frac{1}{20} & \frac{3}{20} \\
\end{array}
\right)_{kl}\tilde {\cal R}_l \,.
\end{eqnarray}
This is the particular case for $N=5$ of the general relation 
Eq.~(\ref{Oodd}),  valid for any $N$.

\section{An explicit example for $N_c=3$}

We illustrate here the construction of the baryon states used in this paper,
given in Eq.~(\ref{MS}), on the example of the $N_{5/2}$ state. This is a 
member of the $70^-$ multiplet for $N_c=3$ with quantum numbers $I=1/2, J=5/2$.  

We start by constructing the 
$MS$ basis of spin-flavor states given in Eq.(\ref{phidef}). They are defined as  
$| \phi_k \rangle= (P_{1k} - 1) | \phi \rangle$, where
the reference state $|\phi\rangle$ is chosen as an eigenstate of $S^2,S_3,I^2,I_3$
\begin{eqnarray}
|\phi\rangle = 
\Big(|q_1\rangle \otimes |i_c = 1 \rangle \Big)^{I=\frac12,\ S=\frac32}_{I_3=+\frac12, \ S_3=+\frac32}
  = \frac{1}{3} \Big( 2 \du \uu \uu - \uu \uu \du - \uu \du \uu \Big)\ . 
\end{eqnarray}
We indicate explicitly the spin and isospin couplings, which were dropped in 
the main text.  
The normalization of $| \phi \rangle$ is chosen so that the basis 
states satisfy Eq.~(\ref{phinorm}). We obtain 
\begin{eqnarray}
| \phi_2 \rangle &=& \uu \du \uu - \du \uu \uu \ , \\
| \phi_3 \rangle &=& \uu \uu \du - \du \uu \uu  \ .
\end{eqnarray}
Applying the total isospin operator  
\begin{eqnarray}
I^2 = \frac{9}{4} + \frac{1}{2} \sum_{i<j}   ( \tau_{i}^3 \tau_{j}^3 +  
2 \tau_{i}^+ \tau_{j}^- + 2 \tau_{i}^- \tau_{j}^+ ) 
\end{eqnarray}
with $\tau^\pm = (\tau^1 \pm i \tau^2)/2$,  where $\vec \tau$ are the Pauli matrices, 
it is easy to check that these states have the same quantum numbers as the
reference state $|\phi\rangle$. 
Combining the spin-flavor and orbital components, our symmetrized state Eq.(\ref{MS}) 
is given by
\begin{eqnarray}
|B \rangle = \sum_{k,l=2}^3 \phi_k \chi_l M_{kl} = \phi_2 \chi_2 + \phi_3 \chi_3 - \frac{1}{2} (\phi_2 \chi_3 + \phi_3 \chi_2 ) 
\end{eqnarray}
with the $MS$ orbital basis wave functions
$
\chi_2  = s p^+ s  - p^+ s s \ ,   
\chi_3  = s s p^+  - p^+ s s  \ ,
$
and reads explicitly 
\begin{eqnarray}
| B  \rangle  &=&
\tdu^+ \uu \uu +  \uu \tdu^+ \uu +  \uu \uu \tdu^+ \nonumber  \\ 
& & - \frac{1}{2} \Big( 
\tuu^+ \uu \du + \tuu^+ \du \uu 
+ \uu \tuu^+ \du + \du \tuu^+ \uu
+ \uu \du \tuu^+ + \du \uu \tuu^+  )\ . \label{bsymm}
\end{eqnarray}
This state is symmetric under all three quarks, and is normalized as 
$\langle B | B  \rangle = 9/2$. 

Next we compare this with the state
constructed in Ref.~\cite{Carlson:1998vx} (denoted as the CCGL state) 
which is given in Eq.~(\ref{CCGLstate}). 
Here we always choose to couple the excited quark 
in the first position, which leads to the explicit form 
\begin{eqnarray}
|(LS)J,J_3;I,I_3\rangle_{CCGL} = |(1,\frac32)\frac52,+\frac{5}{2};\frac12,+\frac{1}{2}\rangle 
 &=&
\frac{1}{\sqrt{6}} \Big(\tuu^+ \uu \du + \tuu^+ \du \uu - 2 \tdu^+ \uu \uu \Big)
\label{n52} \ \ \ 
\end{eqnarray}
where $\tilde q^+$ denotes the quark no.~1 with orbital angular momentum 
$|L=1, L_3=+1\rangle = p^+ $.

The CCGL state Eq.~(\ref{n52}) can
be written as $ |CCGL \rangle = \Phi(SI) \otimes |p^+ s s\rangle $ with
$\Phi(S=\frac32, I=\frac12) = \frac{1}{\sqrt{2 \cdot 3}} \Big(| \phi_2 \rangle + | \phi_3 \rangle \Big)$.
The relation of the symmetrized state Eq.~(\ref{bsymm})
with the CCGL state given in Eq.~(\ref{n52}) is given by
\begin{eqnarray}
\frac{1}{\sqrt{3}} \sum_{i=1}^3 P_{1i} |CCGL \rangle = - \frac{\sqrt{2}}{3} | B  \rangle \ 
\end{eqnarray}
and agrees with Eq.~(\ref{Bstate}).

\section{The proof of Eq.~(\ref{symm2body})}

The matrix elements of a 2-body symmetric operator on the $\phi_i$ basis
can be given in closed form, for any $N$. Using the representation of the
2-body operator $(kl) = O_{kl}$ on the basis of Fock states, these matrix elements are
\begin{eqnarray}\label{(kl)}
& & \langle \phi_i | (1k) |\phi_j\rangle = (I_d-I_e)(\delta_{ik} + \delta_{jk})
+ I_d (1 - \delta_{ik}) (1- \delta_{jk}) \,,\qquad k = 2, \cdots , N\\
& & \langle \phi_i | (kl) |\phi_j\rangle = I_d \delta_{ij} 
(\delta_{ik} + \delta_{jl})
+ I_e (\delta_{ik} \delta_{jl}+ \delta_{il} \delta_{jk}) \,,\qquad
k\neq l = 2, \cdots , N \,.
\end{eqnarray}
We denoted here the two overlap integrals as $I_d = \langle sp |(12)|sp\rangle\,,
I_e = \langle sp |(12)|ps\rangle$.

Using these expressions, the matrix element of operators transforming into
irreps of $S_N$ can be obtained by projecting onto the appropriate irrep.
\begin{eqnarray}
&& \langle \phi_i |O_S|\phi_j \rangle = 
\langle \phi_i |\sum_{1\leq k<l\leq N}(kl) |\phi_j \rangle = 
\langle O_S\rangle (1 + \delta_{ij}) \\
&& \langle \phi_i |O^{MS}_k |\phi_j \rangle = 
\langle O_{MS}\rangle (1 - \delta_{ij}\delta_{ik}) \\
&& \langle \phi_i |O^{E}_{klm} |\phi_j \rangle = 
\langle O_{E}\rangle 
\frac12 [(-\delta_{ik}+\delta_{im})(1+\delta_{jl}) + 
(-\delta_{jk}+\delta_{jm})(1+\delta_{il})]
\end{eqnarray}
where the reduced matrix elements are
\begin{eqnarray}\label{OSred}
&& \langle O_S\rangle = (N-1) I_d - I_e \\
&& \langle O_{MS}\rangle = (N-2) I_d - 2 I_e\\
&& \langle O_E\rangle = 2I_e\,.
\end{eqnarray}

The matrix element of the  interaction Hamiltonian in Eq.~(\ref{symm2body}) 
can be computed as
\begin{eqnarray}\label{LHS}
\langle B | V_{symm} |B\rangle = 
\sum_{1\leq k <l < N} \mbox{Tr }[M(kl)M(kl)] = \frac12 N(N-1) T
\end{eqnarray}
where $M$ is the matrix in Eq.~(11) of the paper, and the matrices $(kl)_{ij}$
are given above in Eqs.~(\ref{(kl)}). In the last step we used the fact that
each term in the sum is the same $T$, such that the sum is simply the number
of terms times $T$.

The result for the matrix element of $V_{symm}$ must be given equivalently by the
equation (\ref{symm2body}) as a sum over the irreps $S,MS,E$
\begin{eqnarray}\label{V3c}
\langle B | V_{symm} |B\rangle = \frac{N^2}{N-1} \Big(
c_S \langle O_S\rangle^2 + c_{MS} \langle O_{MS}\rangle^2 +
c_E \langle O_E\rangle^2 \Big)
\end{eqnarray}
where the reduced matrix elements are given above in Eqs.~(\ref{OSred}) in
terms of the overlap integrals $I_d, I_e$. Comparing the two expressions for
this matrix element allows the determination of the coefficients $c_{S,MS,E}$.

We start by computing the matrix element Eq.~(\ref{LHS}). As mentioned,
each term in the sum is the same, and can be computed as
\begin{eqnarray}
T \equiv \mbox{Tr }[M(1k)M(1k)] = 2\Big(1 + \frac{1}{(N-1)^2}\Big) (I_d^2 + I_e^2)
- \frac{8}{N-1} I_d I_e \,.
\end{eqnarray}
This must be matched by the expression Eq.~(\ref{V3c})
\begin{eqnarray}
& &\langle B | V_{symm} |B\rangle = \frac12 N(N-1) T\\
& & =  
\frac{2N}{(N-1)^2} [(N-1) I_d - I_e]^2 + \frac{N}{N-1} [(N-2)I_d - 2I_e]^2 +
c_E \frac{N^2}{N-1} [2I_e]^2 \nonumber
\end{eqnarray}
Matching the coefficients of $I_d^2, I_d I_e$ and $I_e^2$ on both sides
of this relation, gives the coefficients of the three terms in Eq.~(\ref{V3c})
corresponding to the three irreps of $S_N$
\begin{eqnarray}
c_S = \frac{2}{N(N-1)}\,,\qquad c_{MS} = \frac{1}{N}\,, \qquad 
c_E = \frac{N(N-3)}{4(N-1)}
\end{eqnarray}
This completes the proof of Eq.~(\ref{symm2body}).

\section{The reduced matrix element $\langle O_{MS}\rangle$ }

We present here the details of the derivation of
the reduced matrix element of a $MS$ spin-flavor operator as a 
matrix element on the CCGL state $\Phi(SI)$ with fixed identity of the `excited' quark
(such as e.g. Eq.~(\ref{OMS})).
As explained in Sec.~\ref{Sec:MSstates}, this state is given by 
\begin{eqnarray}
\Phi(SI) = \frac{1}{\sqrt{N(N-1)}}\sum_{k=2}^N \phi_k\,.
\end{eqnarray}

The matrix element of an $MS$ operator on the $MS$ basis states 
is given by the $S_N$  Wigner-Eckart theorem 
\begin{eqnarray}
\langle \phi_i | {\cal O}^{MS}_k |\phi_j \rangle &=& 
\langle {\cal O}^{MS} \rangle (1 - \delta_{ik}\delta_{ij}) \,.
\end{eqnarray}
Summing over the index $k$ of the operator, the matrix element on the $\Phi(SI)$ state is
\begin{eqnarray}\label{B3}
\langle \Phi(SI) |\sum_{k=2}^N {\cal O}^{MS}_k |\Phi(SI) \rangle &=& 
(N-2) \langle {\cal O}^{MS} \rangle \,,
\end{eqnarray}
which can be used to express $ \langle {\cal O}^{MS} \rangle$ as a matrix element on the
CCGL-type spin-flavor state $\Phi(SI)$.

The advantage of taking the sum $\sum_{k=2}^N {\cal O}_k^{MS}$ is that it singles
out the quark no.~1, just as in the state $\Phi(SI)$. An explicit calculation gives
\begin{eqnarray}
\sum_{k=2}^N  {\cal O}_k^{MS} =
\left\{
\begin{array}{cc}
- 2 \sum_{i<j} {\cal O}^s_{ij} + N \sum_{i=2}^N {\cal O}^s_{1i} & 
      \mbox{for symmetric two-body operators} \\
N \sum_{i=2}^N {\cal O}^a_{1i} & \mbox{for antisymmetric two-body operators} \\
\end{array}
\right.
\end{eqnarray}
Using this relation, one finds for example for the spin-spin interaction ${\cal O}_{ij}=
\vec s_i \cdot \vec s_j$
\begin{eqnarray}
\sum_{k=2}^N {\cal O}_k^{MS} = -\vec S\,^2 + N \vec s_1 \cdot \vec S_c + \frac34 N
\,,
\end{eqnarray}
which gives directly Eq.~(\ref{OMS}) after combining it with Eq.~(\ref{B3}).

\end{document}